\documentclass{PoS}

\usepackage{amsmath,amssymb,amsfonts,bbm,verbatim}
\usepackage{arydshln}

\newcommand{\beq}{\begin{equation}}
\newcommand{\eeq}{\end{equation}}
\newcommand{\ba}{\begin{array}}
\newcommand{\ea}{\end{array}}

\title{TASI Lectures on Geometric Tools for String Compactifications}

\ShortTitle{Geometric Tools for String Compactifications}

\author{\speaker{Lara B. Anderson} and {Mohsen Karkheiran}%
       \\
      Virginia Tech, Department of Physics.\\
      E-mail: \email{lara.anderson@vt.edu}, \email{mohsenka@vt.edu}}

\abstract{In this work we provide a self-contained and modern introduction to some of the tools, obstacles and open questions arising in string compactifications. Techniques and current progress are illustrated in the context of smooth heterotic string compactifications to $4$-dimensions. Progress is described on bounding and enumerating possible string backgrounds and their properties. We provide an overview of constructions, partial classifications, and moduli problems associated to Calabi-Yau manifolds and holomorphic bundles over them.}

\FullConference{Theoretical Advanced Study Institute Summer School 2017 "Physics at the Fundamental Frontier"\\
		 4 June - 1 July 2017\\
		 Boulder, Colorado}

\begin{document}

\section{Introduction}\label{intro}
The subject of string compactification -- reducing a 10- or 11-dimensional string theory to a lower dimensional theory by choosing backgrounds of the form $M_{10}=M_{10-n} \times M_{n}^{compact}$, where $M_{n}^{compact}$ is an $n$-dimensional compact space -- has a now more than 30 year history \cite{Candelas:1985en} in string theory and has given rise to an enormous zoo of theories, insights and techniques. Despite a vast literature and substantial progress however, many open questions remain and new methods are being sought to understand more fully how to link the geometry of the compact directions $M_{n}$ with the effective lower dimensional field theory on $M_{10-n}$ that arises from compactification.

In many respects string theory has proven itself to be a natural extension of quantum field theory. But unlike field theory, the rules of string compactifications are not fully understood. That is, for the most part we know how to engineer a consistent quantum field theory with a given particle content or vacuum structure quite explicitly. However, the same questions in string theory -- for example asking to build a theory with the matter and interactions of the Standard Model of particle physics -- are not so clear. The process of string compactifications requires a choice of compact manifold in order to reduce a higher dimensional theory to a lower dimensional one, and the properties of $M_{n}^{compact}$ fix almost all features of the lower dimensional theory. We cannot simply state that the Standard Model arises from string theory, we must engineer a geometry that will yield exactly the required particle content, masses, couplings, etc. The question of "which field theory?" has been replaced with "which geometry?" and in general, the latter is a difficult one that frequently leads us to the cutting edge of modern mathematics.

\begin{figure}
\centering
\includegraphics[width=0.5\textwidth]{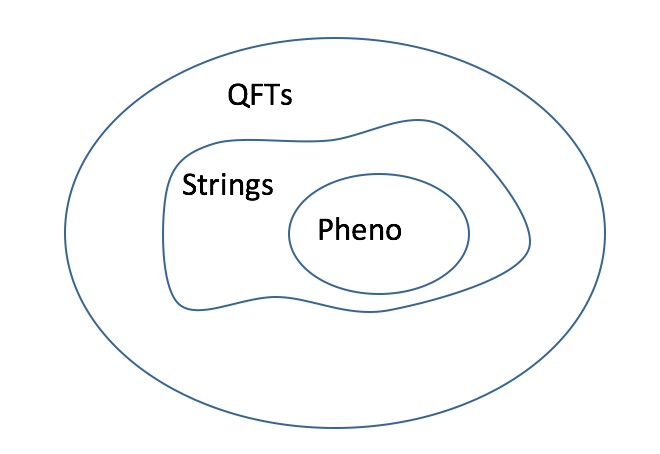}
\caption{\emph{An illustration of the open question of how string effective theories are related to all field theories and those of particular interest (here labeled ``Pheno").}}\label{landscape}
\end{figure}

A question that will motivate us in this lecture series is \emph{what effective theories can arise from string theory?} See Figure \ref{landscape}. Because of the large number of vacua that can occur in string compactifications, there is sometimes a temptation to assume that ``anything is possible" in such a vast landscape. Surely in all the multitude of possible theories that can arise, the one we want (whatever that may be!) is out there somewhere. Or on the negative side, since there is such an abundance of effective theories, perhaps this indicates that string theory has no hope of providing useful new physics. 

We hope to illustrate in these lectures that this viewpoint could lead one to miss important structure. Far from being a framework in which anything goes, the theories that arise from string compactifications can be constrained and intricately related. Caution should also be taken in dealing with the notion of a string landscape (see also the lectures on the Swampland \cite{Brennan:2017rbf} in this School). For instance, the Standard Model is one of an infinite number of quantum field theories, but this isn't really relevant since we understand \emph{how to choose and define} that theory out of all possibilities. Such rules are not fully understood in string theory. For example, despite their vast multiplicity, the famous $10^{500}$ flux vacua \cite{Ashok:2003gk} of Type II compactifications all give rise to effective theories without an electron. If we ask to find theories with such a particle, or perhaps exactly the three families of quarks and leptons in the Standard Model  or a string model of inflation or whatever theory is of interest, the answer may well be that such string vacua are highly constrained or not possible at all.

With these ideas in mind, in this set of lectures we hope to touch on some of the following:

\begin{itemize}
\item Explore the interplay between geometry and effective field theory, known constraints and what is understood about the effective physics.

\item Ways that possible theories arising from string compactifications may be bounded or related (i.e. by string dualities). 

\item We hope (optimistically) to search for patterns, constraints and predictions. Note that some of our motivations include particle phenomenology, cosmology, etc in the context of $4$-dimensional effective field theories. However, for this lecture series we will focus on the basic mechanisms of string compactifications, leaving string phenomenology for another day.

\item We will illustrate tools in the context of $\mathcal{N}=1$, $4$-dimensional effective theories (though many of the tools will apply to theories in other numbers of spacetime dimensions, which you will hear about in other parts of this school).
\end{itemize}

These goals and ideas have been attractive for more than 30 years, so we certainly won't be able to do justice to the literature in this brief space and these remain very hard questions. However, we do hope to report on some areas of progress and several new tools. 

A caveat is also warranted about the level of these lectures. Some people in this audience are already experts, while some have never encountered the mathematics/geometry that arises in string compactifications. It is our aim to try to be pedagogical and reasonably self-contained, while also highlighting new developments and interesting advances too. We have aimed the level of this text for a string theory graduate student who is not currently working on this topic, but would like an intuitive overview of where the open questions lie and \emph{why they are interesting}.

There are a number of useful reviews and lecture courses on string compactifications. We try where possible not to duplicate efforts with some of these classic references, while still making these lectures somewhat self-contained. We recommend to the reader just beginning to study this subject the classic lecture notes on complex geometry by Candelas \cite{Candelas:1987is}, as well as the canonical text by Green, Schwartz and Witten \cite{GSW2} (see Ch. 12). See also \cite{Font:2005td} and for a careful introduction to some of the underlying mathematics used here we recommend \cite{GH,Hartshorne,Huybrechts}.

To begin, we'll illustrate some of these ideas in the context of heterotic string compactifications. The tools we hope to illustrate can be applied much more generally to string compactifications (i.e. Type II theories, F-theory, M-theory compactifications, etc), but we find the heterotic theory to be useful as a simple arena to introduce essential ideas and obstacles. In particular heterotic theories provided the first examples of string compactifications/ dimensional reduction in the literature (see \cite{Candelas:1985en}). They are also appealing in that a great deal of interesting structure can be explored in the context of the perturbative theory --- where there is comparatively more calculational control -- rather than in non-perturbative sectors whose effective physics can frequently be more difficult to obtain. With this in hand, we turn now to the first (in more senses than one!) example of a string compactification.

\section{An Illustration: Smooth Compactifications of the Heterotic String}\label{het_sec}
The heterotic string provides a straightforward arena in which the interplay between geometry and physics can readily be explored. So let's start there and get a feel for things\footnote{Note that part of the content of this Section was based on a joint lecture course that L.A. gave with J. Gray as part of the 2012 Graduate Summer School on String Phenomenology held at the Simons Center for Geometry and Physics. We thank J. Gray for letting us include it here.}.

\subsection{Dimensional Reduction}\label{lagrangian_etc}
At energy levels low with respect to the string scale, we can define the theory by the 10-dimensional action (for simplicity, bosonic part only presented here):

\begin{equation}\label{10D_lag}
S_{10} \simeq \frac{1}{2\kappa_{10}^2} \int_{M_{10}} \sqrt{-g} e^{-2\phi}\left[R + 4(\partial \phi)^2 - \frac{1}{2}H^2 + \frac{\alpha'}{4} tr R^2 - \frac{\alpha'}{4} tr F^2 + \ldots \right] +(\rm{Fermionic Terms})
\end{equation}
where $\frac{\alpha'}{4}=\frac{\kappa_{10}^2}{g^2}$. Here the field content consists of a Yang-Mills multiplet $(A^{A}_{M},\chi^{A})$ (gauge field and gaugino) and supergravity multiplet $(e^{A}_{M}, B_{MN},\phi, \psi_M, \lambda)$ (vielbein, NS 2-form, Dilaton, gravitino, and Dilatino, respectively) where $M,N=1, \ldots 10$ represents the 10-dimensional spacetime indices and $A=1,\ldots N$ is the gauge index. These fields must also satisfy the Bianchi identity
\begin{equation}\label{anom_canc_orig}
dH=\frac{\alpha'}{4}(tr(R \wedge R)- tr(F \wedge F))
\end{equation}
where $H_3 \sim dB_2 - \omega^{YM}_3+\omega^{L}_{3}$ is the field strength associated to the $2$-form. Here $\omega_3$ denotes a Chern-Simons $3$-form $\omega_{3}=A_{a}F^{a} - \frac{1}{3}gf_{abc}A^{a}A^{b}A^{c}$, where $\omega_3^{YM}$ is built from the Yang-Mills connection and field strength and $\omega_3^{L}$ denotes the Chern-Simons $3$-form defined with respect to the spin connection and curvature.

As described in \cite{GSW2}, this action is invariant under the following ${\cal N}=1$ supersymmetry variations of the fermions:\\
\begin{eqnarray} \label{fermion_variation}
\delta\psi_{M} & = & \frac{1}{\kappa}D_{M}\epsilon + \frac{1}{8\sqrt{3}\kappa}e^{-\phi}({\Gamma_M}^{NPQ}-9{\delta^N}_{M}\Gamma^{PQ})\epsilon H_{NPQ} + \ldots \\ 
\delta\chi^{a}  & = &  -\frac{1}{2\surd{2}g}e^{-\frac{\phi}{2}}\Gamma^{MN}{F^a}_{MN}\epsilon +\ldots \label{bundle_eom} \\ 
\delta\lambda & = & -\frac{1}{\surd{2}}(\Gamma\cdot\partial\phi)\epsilon + \frac{1}{4\surd{6}\kappa}e^{-\phi}\Gamma^{MNP}{\epsilon}H_{MNP} +\ldots 
\end{eqnarray}
where $\epsilon$ is 16-component, 10-dimensional (Majorana-Weyl) spinor parameterizing the supersymmetry.

Our first task will be to search for solutions of this theory. To begin, we will try to find solutions that respect some supersymmetry. In particular, we'll consider vacua preserving $\mathcal{N}=1$ SUSY in the 4-dimensional theory. There are many good motivations for such a choice (for example, phenomenological considerations might lead us towards theories that allow chiral fermions or we might be interested in supersymmetry to protect the Higgs) but for now we will take this path because it will allow us to say quite a bit about the structure of the resulting $4$-dimensional theory.

To find such solutions we must check that they respect supersymmetry (i.e. no spontaneous breaking) and look for solutions in $M_4 \times M_6$ with $M_6$ compact. This choice leads to a decomposition of the SUSY parameter $\epsilon$ as
\begin{equation}
\epsilon= \theta \otimes \eta + \theta' \otimes \eta^*
\end{equation}
where $\theta$ is a 4-dimensional spinor and $\eta$ that in the six-dimensional space. In terms of representation theory this is a decomposition of the group
\begin{equation}
SO(1,9) \supset SO(1,3)\times SO(6)
\end{equation}
and its spinor respresentations
\begin{equation}
16 \supset (\textbf{2},\textbf{4})+(\textbf{2}',\textbf{4}')
\end{equation}

Making a choice to preserve $1/4$ of the 10-dimensional supersymmetry means that just one of the four possible $\eta$s is needed. Let's call this $\eta_0$. Then the following fermionic variation must vanish,
\begin{eqnarray} \label{fermion_variation_decompose}
0 &=&\delta_{\eta_0}\psi_{M}  =  \frac{1}{\kappa}D_{M}\theta \otimes \eta_0 + \frac{1}{8\surd{3}\kappa}e^{-\phi}({\Gamma_M}^{NPQ}-9{\delta^N}_{M}\Gamma^{PQ})\theta \otimes \eta_0  H_{NPQ} + \ldots \\ 
0 & =& \delta\chi^{a}   =   -\frac{1}{2\surd{2} g}e^{-\frac{\phi}{2}}\Gamma^{MN}{F^a}_{MN}\theta \otimes \eta_0 +\ldots \\ 
0 &=& \delta\lambda =  -\frac{1}{\surd{2}}(\Gamma\cdot\partial\phi)\theta \otimes \eta_0  + \frac{1}{4\surd{6}g^{2}}e^{-\phi}\Gamma^{MNP}{\theta \otimes \eta_0 }H_{MNP} +\ldots 
\end{eqnarray}

At this point a typical ansatz is to take $M_4$ to be maximally symmetric, and as we described above, and $M_6$ to be compact. It follows then that
\begin{equation}\label{max_symm_constraint}
H_{\mu M N}=0~~~\text{and}~~~\partial_{\mu}\phi=0
\end{equation} 
(in order to make sure $M_4$ has no preferred directions) where $\mu,\nu=1,\ldots 4$ are the coordinate indices on $M_4$. Let $i,j$ be indices running over the coordinates of the $6$-dimensional space. The constraints in (\ref{max_symm_constraint}) leads to, for example the following two conditions
\begin{eqnarray}
\delta_{\eta_0} \psi_{i} &=& D_i \eta_{0}+\frac{1}{8} H_i \eta_0 = 0\\
\delta \psi_{\mu} &=& D_{\mu}^{4D}\theta =0
\end{eqnarray}
From this second equation, we can phrase a constraint as an integrability condition:
\begin{eqnarray}\label{psi_Variation}
\Gamma^{\mu \nu} [D_{\mu}^{4D},D_{\nu}^{4D}]\theta=0, \\
\Rightarrow {R}^{4D}=0
\end{eqnarray}
That is, the curvature over the 4-dimensional spacetime must be Ricci flat and $M^4$ is a Minkowski spacetime (i.e. no other maximally symmetric space is possible. See \cite{Lukas:2010mf, Gray:2012md} for example for other, domain wall type solutions).

This just leaves (\ref{psi_Variation}), as well as
\begin{align}
&\delta \lambda \Rightarrow &(\Gamma.\partial \phi +\frac{1}{12} H_{ijk}\Gamma^{ijk})\eta_0= 0,\\
&\delta \chi^A \Rightarrow &\Gamma^{ij} F^A_{ij}\eta_0 = 0
\end{align}

Now, from (\ref{psi_Variation}) we obtain $(D_i+\frac{1}{8} H_i)$ acting as derivative operator on $\eta_0$ to give zero. This means $\eta_0$ is a nowhere vanishing (since it must parameterize supersymmetry) but not quite covariantly constant spinor ($D + \frac{1}{8}H)$ is nearly a derivative operator, but has been twisted by the $3$-form $H$). The existence of such a spinor guarantees that the compact manifold $M_6$ admits what is called an "$SU(3)$ structure". We refer the reader to \cite{Grana:2005sn} for a nice review of $G$-structure manifolds, but here we'll just provide a little bit of intuition.

\begin{figure}[h]
\centering
\includegraphics[width=0.4\textwidth]{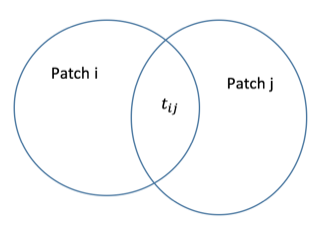}
\caption{\emph{Two coordinate patches on a manifold, including an overlap region on which a transition function, $t_{ij}$ (mapping one set of coordinates to another) is defined.}}\label{transitionfn}
\end{figure}

Where is the $SU(3)$ arising from above? Recall that the spinor of $SO(6)$ is also the fundamental of $SU(4)$ and thus we can write $\eta_0$ as a four-component object. On any given coordinate patch of the manifold (see Appendix \ref{appendix} for a discussion of manifolds and coordinates) the fact that $\eta_0$ is nowhere vanishing means we can find local transformations to put it into the form 
\begin{eqnarray}
\eta_0|_{u_1} = 
\left(\begin{array}{lcr}
0\\
0\\
0\\
\chi
\end{array} \right)
\end{eqnarray}
without loss of generality. But what happens when we move to another coordinate patch? (See Figure \ref{transitionfn} for an illustration of coordinate patches and their overlaps). In a region where both coordinate frames are valid (the overlap region in Figure \ref{transitionfn}) a matrix transformation must rotate the vector from one coordinate frame to another. Let $t\in SU(4)$, be such a transition function, our solution requires that these transition functions must respect this choice of $\eta_0$. 
\begin{eqnarray} 
t_{ij}\left(\begin{array}{lcr}
0\\
0\\
0\\
\chi
\end{array} \right) = 
\left(\begin{array}{lcr}
0\\
0\\
0\\
\chi
\end{array} \right) 
\end{eqnarray}
It follows that the most general matrix to do so is of the form
\begin{eqnarray}
t = 
\left(\begin{array}{cccc}
 &  &  & 0\\
 & SU(3) &  & 0\\
 &  &  & 0\\
0 & 0 & 0 & 1
\end{array} \right)
\end{eqnarray}
More formally, this should be defined in terms of the structure group of the frame bundle\footnote{Which encodes information about a basis of vectors $e_i$ such that $e_i^m e_j^n g_{mn} = \delta_{ij}$ over every point.} becoming reduced from $SO(6)$ to $SU(3)$. See \cite{Grana:2005jc} for a more detailed treatment.

For now we will study the rough properties of these $SU(3)$ structure manifolds. To proceed, it's helpful to build several nowhere vanishing differential forms out of the spinor $\eta_0$ (see e.g. \cite{Gurrieri:2002wz}): \\

$\bullet$ 2-form:  $J_{ij}=-i\eta_0^{\dagger} \Gamma_7 \Gamma_{ij}\eta_0$,

$\bullet$ 3-form: $\Omega =\Omega^+ +i \Omega^-$ \\
\indent with $\Omega^+_{ijk}=-i\eta_0^{\dagger}  \Gamma_{ijk}\eta_0$  and
$\Omega^-_{ijk}=-i\eta_0^{\dagger}\Gamma_7  \Gamma_{ijk}\eta_0$
(where $\Gamma_7$ is the usual product of $\Gamma$-matrices). \\

Using Fierz identities, and suitable normalization, it can be shown that these forms must obey the following relations:
\begin{eqnarray}
J\wedge \Omega &=& 0,\nonumber\\
J\wedge J\wedge J &=& -\frac{3}{4i} \bar{\Omega} \wedge \Omega
\end{eqnarray}

It is a remarkable fact that the possible solutions/G-structures can be classified in terms of $J$ and $\Omega$. The exterior derivatives of these two objects can be described in the following way 
\begin{eqnarray}
dJ &=& - \frac{3}{2} Im(W_1 \bar{\Omega}) + W_4\wedge J +W_3,\nonumber\\
d \Omega &=& W_1 J\wedge J +W_2 \wedge J +W_5 \wedge \Omega
\end{eqnarray}
Where the so-called torsion classes\footnote{To see the relationship of these classes to torsion in General Relativity, see e.g. \cite{Gurrieri:2002wz}, Appendix C.} $W_i$'s parameterizing the solution satisfy the following conditions (which are required to make the decompositions above unique):
\begin{eqnarray}
W_3 \wedge J = W_3 \wedge \Omega &=& 0,\nonumber\\
W_2 \wedge J \wedge J &=& 0.
\end{eqnarray}
These torsion classes \cite{Gurrieri:2002wz} are defined in Table \ref{torsion_classes}.
\begin{table}[h]
\begin{center}
\begin{tabular}{|c|c|c|c|}
\hline
\textit{Torsion Class} & \textit{Interpretation} & $SU(3)$ \textit{Rep} & \textit{Form}\\  \hline
$W_1$& $J\wedge d\Omega$ or $\Omega \wedge dJ$  & \textbf{1}+\textbf{1} & function \\
\hline
$W_2$ & $(d\Omega)_0^{2,2}$ &  \textbf{8}+\textbf{8} & \textit{3-form} \\ \hline
$W_3$ &$(dJ)_0^{2,1}+(dJ)_0^{1,2}$ & \textbf{6}+$\bar{\bf{6}}$ & \textit{3-form} \\ \hline
$W_4$& $J\wedge dJ$ &  \textbf{3}+$\bar{\textbf{3}}$ & \textit{1-form} \\ \hline
$W_5$ & $d \Omega^{3,1}$ & \textbf{3}+$\bar{\textbf{3}}$ & \textit{1-form} \\
\hline
\end{tabular}
\caption{\emph{Summary of torsion classes parameterizing $SU(3)$ structure manifolds \cite{Gurrieri:2002wz}.}}\label{torsion_classes}
\end{center}
\end{table}

So to solve the heterotic compactification in general (with $M_4$ maximally symmetric) we need a solution to the so-called ``Strominger System" \cite{Superstrings with torsion,Hull:1986kz} (see also \cite{LopesCardoso:2002vpf} for the set up of this system in modern language)
\begin{eqnarray}\label{Strominger_System}
W_1=W_2 &=& 0,\nonumber\\
W_4=\frac{1}{2} W_5 &=& exact,\nonumber\\
W_3 &=& free.
\end{eqnarray}
We have illustrated this here in the heterotic theory, but it should be noted that for any string compactification on an $SU(3)$-structure manifold, properties of the $4$-dimensional theory an be written in terms of these torsion classes. As a result, this is a useful description of a solution. 

First we must face the question: how can one find the allowed $J$'s and $\Omega$'s? There are some options, 
\begin{itemize}
\item Solve for possible $\eta_0$'s then plug in. This is hard.
\item Restructure SUSY equations of motion in terms of forms, then look at constraints in terms of torsion classes instead. This is still hard, but some progress is possible.
\end{itemize}
With this in mind, let's return to the Strominger system (\ref{Strominger_System}) with a bit more detail:
\begin{eqnarray}
W_1=W_2 &=& 0,\nonumber\\
W_4=\frac{1}{2} W_5 &=& d\phi,\nonumber\\
W_3=free.
\end{eqnarray}
$W_3$ is free because it can be balanced with a correct choice of flux $H \ne 0$. Returning then to the point above that torsion classes can determine the $4$-dimensional theory, note that 
for example, the Gukov-Vafa-Witten superpotential of the ${\cal N}=1$ theory in $4$-dimensions is given by 
\begin{equation}
W \sim \int \Omega \wedge (H+i dJ),
\end{equation}
where in the Strominger System, after solving the equations of motion, $H \sim W_3$, and $J\sim W_1$. So here torsion classes determine the $4$-dimensional superpotential (we'll return to the superpotential in more detail in Section \ref{bundle_sec}).

One other important observation can be made here about the geometry of the Strominger system. The condition that $W_1=W_2 =0$ guarantees that $M_6$ is actually a complex manifold (see Appendix \ref{appendix} for definitions). All solutions to the Strominger system are complex manifolds and from now on we will refer to such a space by its complex dimension as $X_3$. In some special cases (that we will explore in the next Section) it may also be true that $dJ=0$. This is the case of so-called K\"ahler manifolds. Let's now turn our attention to looking for solutions to the Strominger system.

\section{Manifolds}
Our search for solutions to equations of motion arising in a heterotic compactification have lead us to constrained G-structure manifolds. It is natural to ask, what solutions are known? \\

\noindent \textbf{Example 1: Calabi-Yau Solutions} \\
\noindent The simplest possible solution is clearly that which sets 
\begin{equation}
W_1=W_2=W_3=W_4=W_5=0
\end{equation}
It follows that $H=0$, $\phi$ will be constant, and $\eta_0$ will be nowhere vanishing and in this case, \emph{covariantly constant} since $D_i \eta_0 =0$. By Berger's classification \cite{berger} the existence of such a covariantly constant spinor  implies that $X$ is no longer just an $SU(3)$ structure manifold, in this case it will have $SU(3)$ \emph{holonomy} (see e.g. \cite{GSW2}). The supersymmetry variations thus require the manifold to be Ricci flat to first order in $\alpha'$. 
Since $dJ=0$ in this case,

\vspace{7pt}

\noindent \textbf{Definition:} A Calabi-Yau manifold is a compact, complex manifold which admits a (unique) Ricci-flat Kahler metric $g_{a \bar{b}}$ (in complex coordinates) with $SU(N)$ holonomy.

\vspace{7pt}
\noindent Since $X_3$ is Ricci flat, it follows that the first Chern class, $c_1(X)$ (see Appendix \ref{appendix} for definitions) which is proportional to $tr(R)$ vanishes\footnote{Getting a bit ahead of ourselves, in general for the Strominger System, the existence of the nowhere vanishing 3-form $\Omega \in H^0(X, \Lambda^3 TX^{*})$ implies that $\Lambda^3 TX^{*}$ is the trivial line bundle ${\cal O}_X$ and $c_1(TX)=0$ for all such heterotic solutions.}. Yau's proof \cite{Yau1,Yau2} of Calabi's conjecture \cite{Calabi} proved that any K\"ahler manifold with $c_1(X)=0$ admits a Ricci-flat metric.

Such solutions appeared in \cite{Candelas:1985en} and of course have dominated the literature ever since. As we'll discuss further in subsequent sections, about half a billion distinct Calabi-Yau manifolds are known (see \cite{KS} for example). However, despite the fact that there are numerous Calabi-Yau solutions, it is clear from (\ref{Strominger_System}) that they are by no means the most general class of solutions. However, once we begin to have some non-vanishing torsion classes, the known examples become far more scarce.  Let's turn to one class of such examples now. \\

\begin{figure}[h]\label{FuYau}
\centering
\includegraphics[width=0.5\textwidth]{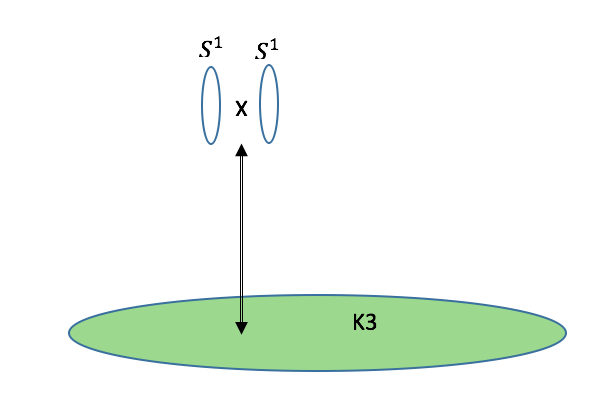}
\caption{\emph{An illustration of a non-K\"ahler, $SU(3)$-structure manifold constructed as a non-trivial fibration of $S^1 \times S^1$ over a $K3$ surface \cite{Fu:2006vj,Goldstein:2002pg}.}}
\end{figure}

\noindent \textbf{Example 2: The Fu-Yau construction} \\
The first class of Strominger system solutions with non-trivial torsion arose from two important developments. Goldstein and Prokushkin formulated a construction of simple $SU(3)$-structure manifolds in \cite{Goldstein:2002pg} . Following on from this work, Fu and Yau introduced suitable gauge fields (i.e. a vector bundle) to solve the equations of motion, and Bianchi identity \cite{Fu:2006vj}. In this class of solutions $X$ is defined via a non-trivial twisting (i.e. fibration) of $S^1 \times S^1$ over a $K3$-surface (a CY $2$-fold) as illustrated in Figure \ref{FuYau}. The full solution yields a 2-parameter family. However, unfortunately, it has been shown that all such manifolds have small cycles (i.e. regions with very small volumes) \cite{Cyrier:2006pp}. Therefore the supergravity limit may not be a valid solution for the string theory in this situation.

In a similar spirit, another systematic construction of solutions to the Strominger system recently been proposed by Teng Fei and collaborators \cite{Fei:2017ctw} in which $X$ is a fibration of a hyperk\"ahler 4-manifold (either a $T^4$ or a $K3$-surface) over a Riemann surface. In this case, an infinite family of solutions, with distinct topology for $X$ has been constructed. However, once again the existence of small cycles may prove problematic (as well as the structure group of the gauge bundles which may not embed into $E_8$).

And that's it! These examples are essentially the only systematic constructions of heterotic Strominger solutions known\footnote{There are also a handful of solutions leading to $\mathcal{N}=\frac{1}{2}$ in 4-dimensions: domain wall, half-flat solutions, etc.}. It is natural to wonder why it is so hard to find solutions in general. The answer lies in the difference between obstacles in differential geometry and algebraic geometry.

\vspace{7pt}

\noindent \textbf{Observation:} The only projective varieties of $SU(3)$ structure are Calabi-Yau manifolds. This follows from the fact that algebraic varieties are complex, K\"ahler manifolds. Once the condition of $SU(3)$ structure is imposed, which leads to the vanishing first Chern class, this leaves us with a Calabi-Yau manifold.

\vspace{7pt}

\subsection{A brief overview of Calabi-Yau manifolds}\label{cy_overview}
We'll begin by providing a current overview of some recent results and investigations into Calabi-Yau manifolds. This is not just motivated by the fact that these provide the easiest class of solutions to the Strominger system. Instead, as we will see in the following sections, this is an arena where we are just beginning (after a century of progress in algebraic geometry, 30 years in string theory, and a handful of Fields Medals) to have the necessary mathematical toolkit available to extract the physics that we're interested in from string compactification! We should bear in mind though that this class of solutions is likely just the tip of the SU(3)-structure iceberg\footnote{Someone once noted that classifying differential equations into linear vs. non-linear was like classifying the universe into bananas vs. non-bananas -- true but not the most enlightening distinction! The same could be said here for solutions with and without torsion.}. 

Even within this class of ``simple" solutions, we will find that difficulties are still plentiful. It is important to recall that Yau's theorem \cite{Yau1,Yau2} guarantees the existence of a unique Ricci flat metric on each manifold of $SU(3)$ holonomy, but no Calabi-Yau (Ricci flat) metrics are explicitly known in general cases. This existence proof has been essential to progress in string compactifications and in particular, made it possible to study large numbers of string vacua (Calabi-Yau backgrounds). In contrast, there is as yet no analog of Yau's theorem for manifolds of $G_2$ holonomy. As a result, explicit examples of M-theory compactifications on $G_2$ manifolds have been much harder to construct. It is only recently that some systematic families have been constructed (see e.g. \cite{Halverson:2014tya,Braun:2018fdp} and \cite{Joyce, Kovalev1,Kovalev2,CHNP1,CHNP2,CHNP3}). 

Even with such a powerful existence proof in the case of Calabi-Yau manifolds, the lack of explicit metrics has still formed a major obstacle. In a string compactification, having an explicit, functional form for the metric would make determining the effective theory much more straightforward. In any dimensional reduction, we would begin with a 10-dimensional lagrangian of the schematic form:
\begin{eqnarray}
S \sim \int_{{X}_{10}} d^{10}x \left( \sqrt{-G} {R}-\partial_{\mu} \mathbb{\phi} \partial^{\mu} \mathbb{\phi}+\dots \right)  \qquad {X}_{10} = {R}^{1,3} \times {M}_6 
\end{eqnarray}
If the metric were known explicitly, it would be possible in principle to just "integrate out" the dependence on the compact directions, ${M}_6$. For this reason (amongst many others), there are still active efforts to explicitly determine Calabi-Yau metrics, including numeric approaches based on the Donaldson algorithm \cite{Donaldson1,Donaldson2,Donaldson3,Tian,Douglas:2006rr,Braun:2007sn} and energy minimization \cite{Headrick:2009jz}.

Without an explicit Calabi-Yau metric, however, we are forced to proceed anyway and try to describe the effective theory that arises from compactification. Fortunately, quite a lot can be said about the form of the metric in a Calabi-Yau manifold. We refer the reader to the well-known summaries of this in \cite{GSW2,Candelas:1987is}. 

The short summary is that a metric on a Calabi-Yau n-fold is linked to two important geometric properties/structures:

\begin{enumerate}
\item \textbf{Complex structure:} Since $W_1=W_2=0$, we get  $J^i_j=g^{ik}J_{kj}= g^{ik} (-i{\eta_0}^{\dagger} \Gamma_7 \Gamma_{kj} \eta_0)$ as before. 

Therefore there is a $J^i_j $ which satisfies $J^2=-\mathbb{I}$ (as well as an integrability condition -- the vanishing of the Niejenhuis tensor, see \cite{Candelas:1987is}). Thus, $M_6$ is a complex manifold.

\item \textbf{K\"ahler structure:} With $dJ=0$, $J$ is a 2-form called a "Kahler form". This leads to $ds^2=g_{a\bar{b}} dz^a dz^{\bar{b}}$, in complex coordinates, with $g_{a\bar{b}}=\partial_a \partial_{\bar{b}} f$. The function $f$ is called a "Kahler potential".
\end{enumerate}

To learn more about why these structures arise, we suggest \cite{Gurrieri:2002wz, Hubsch:1992nu,GSW2}. Much more can be said about complex and K\"ahler manifolds and the structure of a Calab-Yau metric. All of the above properties are important, but they are covered in many places and we will not focus in detail on them here. Instead we'll just provide a simple overview without derivations.

The K\"ahler and complex structures are particularly important because they lead to classification of Calabi-Yau metric moduli. In complex coordinates, $g_{a\bar{b}}$ is the only non-trivial component of metric. Therefore any infinitesimal deformation can be decomposed into one of two index types

\begin{eqnarray}
g\rightarrow g_{(0)}+\delta g \Rightarrow \delta g \Big\lbrace^{\textit{K\"ahler fluctuations}}_{\textit{Complex structure fluctuations}}
\end{eqnarray}
One of these can be simply illustrated and we begin with that result
\begin{itemize}
\item \textbf{Complex structure fluctuations:}  $J$  is closed therefore under small fluctation $d(J+\delta J)$ must still be equal to zero. This implies schematically that
\begin{eqnarray}
J^i_j\rightarrow J^i_j +\delta J^i_j, \quad &\Rightarrow &\delta J^a_b = \delta J^{\bar{a}}_{\bar{b}}=0, \quad \bar{\partial} \delta J^{\bar{a}}_b =0. \\
&\Rightarrow &\delta J^{\bar{a}}_b \in H^1 (TX)=H^{2,1}(X).
\end{eqnarray}
(more precisely, the fact that $\delta J^{\bar{a}}_b$ is closed together with the vanishing of the Niejenhuis tensor gives the above result). Here we have used the notation of cohomology (e.g. $H^1(TX)$) and Hodge numbers (e.g. $H^{p,q}(X)$). We refer the reader unfamiliar with these notions to Appendix \ref{appendix}.

\item \textbf{K\"ahler structure fluctuations} A similar deformation argument can be used to show $\delta J_{a\bar{b}} \in H^{1}(TX^*) = H^{1,1}(X)$.
\end{itemize}
Intuitively, K\"ahler fluctuation corresponds to the ``volume fluctuations" of the manifold\footnote{Note that there can be multiple K\"ahler modes and hence many independent sub-spaces or ``cycles" within the geometry with different volume moduli.}, and complex structure fluctuations corresponds the "shape fluctuations".

It is worth noting briefly for a moment that the word moduli has two different meanings. From the point of view of geometry, metric moduli are simply fluctuations, $\delta g$, which preserve the defining equations of motion (i.e. Ricci-flatness in this case). In contrast, from the point of view of the 4-dimensional effective theory arising from compactification, ``moduli" correspond to flat directions of the potential. These degrees of freedom appear in many aspects of the largrangian of the 4-dimensional theory, from couplings to potentials, and can frequently be unphysical massless scalars. This has led to the vast subject of moduli stabilization within string theory. We will see in later sections that these two notions of moduli (i.e. physical/geometrical) are in fact, the same. 

Before we begin to construct Calabi-Yau manifolds in detail, it should be understood that in order to work out the effective physics from a string compactification, we in fact need much more information than just the metric and its moduli. For example, below is a non-exhaustive list of information about a Calabi-Yau manifold that a string phenomenologist would like to have in hand in order to study the effective theory:
\begin{itemize}
\item Information on "sub-structure" of the manifold. This includes cycles that can be wrapped by branes to produce important non-perturbative physics. In particular, for a Calabi-Yau threefold:
\begin{itemize}
\item Divisors (codimension 1 sub-varieties)
\item Curves (codimension 2 sub-varieties)
\item Special Lagrangian cycles or other non-complex subspaces in $M_6$.
\end{itemize}
\item Symmetries of the manifold? Calabi-Yau manifolds do not admit continuous isometries\footnote{Since $H^0(X,TX)=0$.}, but discrete symmetries can and do arise, and can play an important physical role. Applications of such symmetries include
\begin{itemize}
\item The existence of discrete Wilson lines (i.e. gauge fields for which $A_{\mu}\ne 0$ but $\mathcal{F}_{\mu \nu}=0$, leading to Aharanov-Bohm type effects in the effective theories \cite{GSW2}).
\item Orbifolds and Orientifolds (i.e. related to singular Calabi-Yau geometries and non-perturbative physics). 
\item Flavor/family symmetries/R-symmetry.
\end{itemize}
\item The existence of differential forms. For example, the type of gauge fields, 2-forms (or more generally, n-forms) that can arise on a Calabi-Yau manifold will dramatically effect the type of string backgrounds that are possible. Mathematically this corresponds to the computation of various sheaf and bundle valued cohomology groups. At higher order, interactions in the potential (e.g. Yukawa couplings) can be computed as trilinear couplings of such forms (i.e. Yoneda pairings, see for example \cite{Berglund:1995yu,Anderson:2009ge,Anderson:2010tc}). We'll touch more on these notions in Section \ref{bundle_sec} and in Appendix \ref{appendix}.
\end{itemize}

\subsection{Building Calabi-Yau manifolds}\label{manifold_building}
For now though, we must proceed one step at a time. Having just discovered that we Calabi-Yau manifolds can provide a set of background solutions to our equations of motion, we would like to ask the following questions:

\begin{enumerate}
\item How can we build Calabi-Yau manifolds explicitly?
\item How many are there?
\end{enumerate}

Let's start with the first question and take a bird's eye view of the past few decades of progress in geometry. Basically every known Calabi-Yau manifold that has been constructed in the literature has used the following logic to construct examples: Begin with some simple geometry, then use a simple modification of it to build more complicated geometry, in this case a Calabi-Yau manifold. Let's be a little more precise by what we mean by "start simple".

\textbf{Three main approaches:} (almost all known Calabi-Yau manifolds onstructed using one of these methods)

\begin{enumerate}
\item \textbf{Algebraic Varieties:} Let $X$ denote $M_6$ as a complex manifold. 

\textbf{Idea:} We embed $X$ in a simple (complex) geometry $\mathcal{A}$ as the zero locus of a set of polynomials \begin{equation}
p_1(x_i),p_2(x_i),\dots ,p_n(x_i)=0~. \label{poly_def}
\end{equation}
Where $p_j(x_i)$'s are polynomials in complex coordinates, $x_i$ of $\mathcal{A}$.\\
Here the topological data of $X$ determined by the topology of $\mathcal{A}$ and the form of the polynomial expressions in (\ref{poly_def}). The very simplest example of such a manifold is the famous quintic hypersurface in $\mathbb{P}^4$:
\begin{equation}
\mathbf{P}^4[5], \qquad p=x_0^5 +x_1^5 \dots + x_4^5=0
\end{equation}
(we'll come back and explain this notation properly in Section \ref{alg_sec}).
\item \textbf{Fibrations:} A manifold, $X$, is a fibration and denoted $\pi:X \to B$ where $\pi$ is a surjective morphism.\\
\textbf{Idea:} Just like stacking building blocks, the idea here is to stack or "fiber" one simple geometry over every point of another.\\
\textbf{Definition of fibration:} A continuous mapping 

\[\pi : X\rightarrow B,\]

\noindent such that for (almost all) points $b,b'\in B$, 
\[\pi^{-1}(b) \sim \pi^{-1}(b')~. \]
$B$ is referred to as the "base" of the fibration, while $F=\pi^{-1}(b)$ is the "fiber". Frequently this is denoted $F \rightarrow X \rightarrow B$ for the inclusion of the fiber in $X$ and the mapping of $X$ onto the base. The latter condition above is called "Homotopy lifting" (i.e that most fibers are homotopic, i.e. the same). Most readers will have already come across a number of fibrations in their mathematical education, though they may not have realized it. For example, $\mathbb{R}^2$ is a $\mathbb{R}$-fibration over $\mathbb{R}$. That is over each point in the horizontal axis there exists a copy of the real line. Thus, both base and fiber are $\mathbb{R}$ here (see Figure \ref{R2asfiber}).

\begin{figure}[h]
\centering
\includegraphics[width=0.4\textwidth]{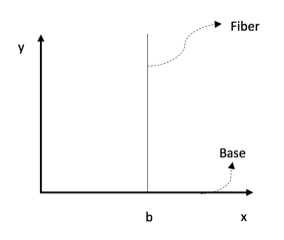}
\caption{\emph{$\mathbb{R}^2$ forms the simplest example of a line fibered over a line.}}\label{R2asfiber}
\end{figure}

Another very simple example is the cylinder, which can be viewed as a line or interval fiber over $S^1$ (depending on whether the cylinder is infinite or finite). See Figure \ref{Cycliderasfiber}

\begin{figure}
\centering
\parbox{5cm}{
\includegraphics[width=5cm]{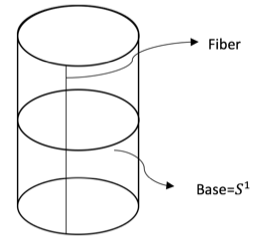}\label{Cycliderasfiber}
\caption{\emph{A line can also be fibered over a compact space (here an $S^1$) as happens in the cylinder.}}
\label{fig:2figsA}}
\qquad
\begin{minipage}{5cm}
\includegraphics[width=5cm]{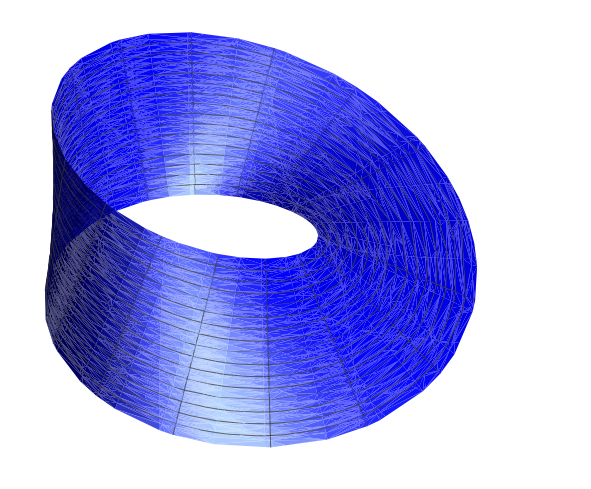}\label{mobiusasfiber}
\caption{\emph{The M\"obius strip is a non-trivial fibration of an interval over $S^1$.}}
\label{fig:2figsB}
\end{minipage}
\end{figure}

\item \textbf{Surgery:} Blow-ups, resolutions, birational correspondences, etc. \\
\textbf{Idea:} The idea is to start with a simple, known space, $X_{easy}$, and locally patch another simple space, $Y_{easy}$, to make a more complicated space, $X'$. 

An example of this is a so-called small resolution, which is a mapping $f: X' \to X_{easy}$ such that for all $r>0$, the space of points in $x \in X_{easy}$ where $f^{-1}(x)$ has dimension $r$ is of co-dimension greater than $2r$. Intuitively this is the statement that $X'$ and $X_{easy}$ look the same almost everywhere, but differ over special, higher-codimensional loci.

\begin{figure}[h]
\centering
\includegraphics[width=0.7\textwidth]{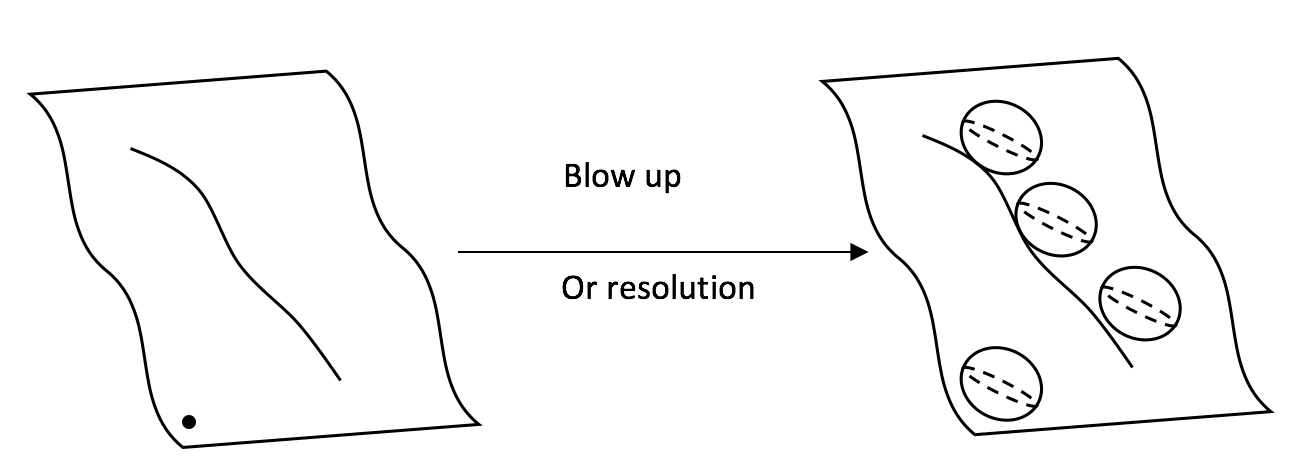}
\caption{\emph{An illustration of the process of blowing up a geometry or resolving a singularity. These are related to the mathematical notion of birational geometry. See e.g. \cite{hacon}.}}\label{Surgery}
\end{figure}
An example of this that we'll see in much more detail in Section \ref{conifold_sec} is the so-called conifold transitions in string theory \cite{Candelas:1989ug}. Very briefly the idea is to start with a smooth Calabi-Yau manifold, deform it until it becomes a singular Calabi-Yau variety and then patch in a new space to "fix the singularity", resulting in a new, smooth Calabi-Yau manifold. 
\end{enumerate}

Remarkably, the vast majority of all Calabi-Yau manifolds constructed have been built using one of the approaches
given above\footnote{An important exception which we won't discuss in detail here is the process of quotienting a Calabi-Yau threefold by a freely acting discrete automorphism, $\Gamma$ to produce $\tilde{X}=X/\Gamma$, a new, non-simply connected Calabi-Yau threefold. Intuitively, this process of quotienting is akin to geometric "origami", folding one space on top of itself to produce another. See \cite{GSW2,Anderson:2009mh} for more details.}. 

\subsection{Algebraic approaches to CY Geometry}\label{alg_sec}
We'll explore each of these methods of building manifolds in turn, beginning with the most intuitive, algebraic constructions. From the description above, we need to begin with an easy complex ambient space in which to define a set of polynomial equation. Perhaps the easiest complex space that comes to mind is $\mathbb{C}^n$, but we recall that we're after compact manifolds. The process of building compact algebraic manifolds is considerably easier if the ambient space we begin with is compact. Fortunately, there is an easy way to turn $\mathbb{C}^n$ into a compact space. Suppose we begin with $\mathbb{C}^{n+1}$ with holomorphic coordinates, $z_i,$ and identified those coordinates up to a scale in the following way
\begin{equation}\label{pn_coords}
\left( z_0,z_1,\dots , z_n \right) \sim \lambda \left( z_0,z_1,\dots , z_n \right), \quad \lambda \ne 0. 
\end{equation}
It is clear that the identification using the overall scale, $\lambda$, prevents any set of points from running away to infinity. This easy space is called ``complex projective space", $\mathbb{CP}^n$ (frequently this is abbreviated to $\mathbb{P}^n$ in contexts where the complex nature of the space is taken for granted and we'll follow that convention here). The $z_i$ are referred to as homogeneous coordinates. The simplest example is the compact, complex curve $\mathbb{P}^1$ which as a real manifold is simply  $S^2$ (see e.g. \cite{Okonek} for why this is the case).

Let us now delve a little more deeply into possible algebraic constructions of Calabi-Yau manifolds and try to define such a geometry as the solution to a set of polynomial equations in $\mathbb{P}^n$. We'll begin with a single hypersurface in an ambient projective space, defined by one holomorphic equation. Define $X_{CY}$ as the zero locus of a polynomial
\begin{equation}\label{hypersurface}
p(z^i)=0
\end{equation}
of homogeneous degree m in the holomorphic projective coordinates (\ref{pn_coords}) of $\mathbb{P}^n$. A common convention is to denote this manifold by a shorthand notation that encapsulates the ambient space and the degree of the equation: $\mathbb{P}^n[m]$. It is important to note that the conditions that the polynomial be homogeneous and holomorphic are crucial for us, otherwise we will not get a well-defined complex manifold (note: if the equation was not homogenous in the projective coordinates, its zero locus would not even be well defined under the scaling relation of (\ref{pn_coords})).

Now, we must ask under what conditions can this projective hypersurface be a Calabi-Yau variety? According to Yau's theorem, we need a K\"ahler manifold with vanishing first Chern class. It can be shown that any complex submanifold of a K\"ahler manifold (defined via holomorphic equations) is K\"ahler \cite{GSW2}, however the condition on the curvature given by $c_1(TX)=0$ is less obvious (see Appendix \ref{appendix} for definitions of Chern classes, etc).

For this latter condition, it should be recalled that $\mathbb{P}^n$ comes equipped with a natural K\"ahler metric, called the \textit{Fubini-Study} metric: 
\begin{equation}
g_{a\bar{b}}=\partial_{a}\partial_{\bar{b}} \log (|z_0|^2+\dots + |z_n|^2)
\end{equation}
which be restricted to the ${p=0}$ hypersurface. Note this restricted metric will not be Ricci-flat! However, it's a good homework problem for any over-achieving student in a General Relativity course to show that the Ricci $2$-Form has the following form:
\begin{equation}\label{cy_metric_restriction}
R_{FS}|_X = ((n+1)-m)g_{FS}+\textit{total derivatves}
\end{equation}
(see \cite{Font:2005td} for details). So if $m=n+1$ then $tr(R)=0$, and this means $c_1(TX)=0$. Therefore if the degree condition is met, $X$ is a Calabi-Yau manifold! The Ricci-flat metric will not be the one we obtained above, but it will share the topological property (invariant under metric deformations) that $tr(R)=0$. 

If we are interested in building a Calabi-Yau threefold in this manner in the context of a string compactification, we are immediately led to the famous quintic hypersurface -- $\mathbb{P}^4[5]$ -- defined by any homogeneous degree $5$ polynomial in the coordinates of $\mathbb{P}^4$, for example the so-called "Fermat" quintic with 
\begin{equation}\label{fermat}
p=z_0^5+\dots + z_4^5=0
\end{equation}
 is a Calabi-Yau manifold. But this is far from the only such defining equation that could be written down. It is a straightforward problem in combinatorics to show that there are $5+4 \choose 4$$=126$ degree $5$ monomials in the homogeneous coordinates of $\mathbb{P}^4$. But not all of these lead to different defining equations. The $PGL(5, \mathbb{C})$ action on $\mathbb{P}^4$ generates $(25-1)$ coordinate redefinitions. Moreover, there is one overall scale that clearly doesn't effect an equation like (\ref{hypersurface}) or (\ref{fermat}). This leaves $126-(25-1)-1=101$ distinct coefficients that do matter. By varying the coefficients of the defining equation in this way, we obtain a different manifold for each choice. This of course, is the complex structure moduli of the threefold visible through this algebraic description\footnote{Note that in a given algebraic description it is not always true that all the complex structure moduli are visible in this way. We'll return to so-called ``non-polynomial" deformations in Section \ref{gcicys}.}. One family choice of degree and ambient space (here $\mathbb{P}^4[5]$) leads to a parametric family of manifolds, all of which can be smoothly deformed into one another by varying the the complex structure. Here there is precisely one further metric modulus, namely the overall volume of the quintic hypersurface in $\mathbb{P}^4$. This is the K\"ahler modulus.
 
 The results for a single hypersurface in a projective space can be easily generalized to complete intersections of polynomial equations $p_j(z_i)=0$ \cite{Candelas:1987kf}. The analog of (\ref{cy_metric_restriction}) in this case is that the zero locus of a set of polynomial equations, $p_j$ of degree $m_i$ in $\mathbb{P}^n$ will be Calabi-Yau when
\begin{equation}\label{partition_prob}
n+1=\sum m_i
\end{equation}
Using this trick, it is clear that we can start with a bigger ambient space -- say, $\mathbb{P}^5$ -- and use \emph{two} defining equations to define a Calabi-Yau threefold. In this case that brings us to $\mathbb{P}^5[3,3]$ or $\mathbb{P}^5[4,2]$. But how many more can we build in this way?

Naively, it seems we could construct infinitely many such manifolds, by letting $n$ become arbitrarily large and adding more and more equations. However, in order to satisfy (\ref{partition_prob}) is clear that we must find a length $n-3$ partition of $n+1$. Above, $\{3,3\}$ and $\{4,2\}$ are two such partitions when $n=5$. But there is another -- what about $\mathbb{P}^5[5,1]$? This is the zero-locus of two equations 
\begin{equation}
p_5(z_i)=0 ~~~\rm{and}~~~p_1(z_i)=0
\end{equation}
Considering the linear, $p_1(z_i)=a_iz_i=0$, it is clear that we can always use this to solve for one of the homogeneous coordinates $z_i$ in terms of the others. This has the simple effect of reducing the number of homogeneous coordinates by one, but keeping the scaling relation of (\ref{pn_coords}). Thus, it is just reducing $\mathbb{P}^n$ to $\mathbb{P}^{n-1}$ (i.e. $\mathbb{P}^n[1]=\mathbb{P}^{n-1}$). So this last choice simply brings us back to the quintic, i.e. $\mathbb{P}^5[5,1]=\mathbb{P}^4[5]$.

This observation bounds the number of Calabi-Yau manifolds we can define in this way rather abruptly. There are in fact only $5$ manifolds (referred to as ``cyclic" Calabi-Yau manifolds). In addition to those above, we add $\mathbb{P}^6[2,2,3]$ and $\mathbb{P}^7[2,2,2,2]$ to complete our set (once we try to define 5 equations in $\mathbb{P}^8$ we are forced to include linear degrees in our partition).

This is not the only way we can build CY manifolds of this type however. The construction above can be readily generalized in several ways. The first we can consider is to take an ambient space which consists of \emph{products} of the simple projective spaces we defined above. To build a Calabi-Yau 3-fold in a product of $p$ projective spaces, $\mathbb{P}^{n_1} \times \ldots \mathbb{P}^{n_p}$, we need
\begin{equation}\label{ci}
\sum_{r=1}^p n_r - K = 3 \ .
\end{equation}
where $K$ is the number of defining equations, $p_j(z)=0$ in the complete intersection. Each of the defining homogeneous polynomials $p_j$ can be characterized by its multi-degree ${\bf m}_j=(m_j^1,\ldots , m_j^m)$, where $m_j^r$ specifies the degree of $p_j$ in the homogeneous coordinates ${\bf z}^{(r)}$ of the factor $\mathbb{P}^{n_r}$ in $\mathcal{A}$.  A convenient way to encode this information is by a {\it configuration matrix}
\begin{equation}\label{cy-config}
\left[\begin{array}{c|cccc}
\mathbb{P}^{n_1} & m_{1}^{1} & m_{2}^{1} & \ldots & m_{K}^{1} \\
\mathbb{P}^{n_2} & m_{1}^{2} & m_{2}^{2} & \ldots & m_{K}^{2} \\
\vdots & \vdots & \vdots & \ddots & \vdots \\
\mathbb{P}^{n_p} & m_{1}^{p} & m_{2}^{p} & \ldots & m_{K}^{p} \\
\end{array}\right]\; .
\end{equation}
Note that the $j^{\rm th}$ column of this matrix contains the multi-degree of the polynomial $p_j$.
In order that the resulting manifold be Calabi-Yau, the condition
\begin{equation}\label{cy-deg}
\sum_{j=1}^K m^{r}_{j} = n_r + 1 \qquad \forall r=1, \ldots, p
\end{equation}
needs to imposed, analogously to (\ref{cy_metric_restriction}) and (\ref{partition_prob}) so that $c_1(TX)$ vanishes.

For example, instead of $\mathbb{P}^4$ we can consider a hypersurface in the simple, 4-complex dimensional ambient space $\mathcal{A}=\mathbb{P}^1 \times \mathbb{P}^3$
\begin{eqnarray}
&& \begin{array}{c}
\mathbb{P}_x^1 \\
\mathbb{P}_y^3
\end{array} \left[  \begin{array}{c}
2 \\
4
\end{array} \right]\\
p&=&x_0^2y_0^4+x_0 x_1 y_0^2 y_1 y_2 +\dots =0
\end{eqnarray}
The dataset of manifolds (sometimes referred to somewhat imprecisely as the ``CICY threefolds", where CICY stands for "complete intersection Calabi-Yau") that can be constructed in products of ordinary projective spaces consists of $7890$ configuration matrices of the form (\ref{cy-config}) and was classified by Candelas, et al \cite{Candelas:1987kf} in a remarkable feat of 1980s computing at CERN\footnote{Even with modern computers, a naive integer partition problem like the one we did above to generate CY manifolds in a single $\mathbb{P}^n$ becomes astonishingly slow if implemented directly for products of projective spaces. Thus, a number of clever subtleties went into proving the set of CICY threefolds was finite and in finding a minimal number of configuration matrices to represent it. See \cite{Candelas:1987kf} for details.}. This formed the first such systematic construction in the literature and as we will see in later sections, set the stage for much important work since! We'll not look at them here, but it should be noted that the Calabi-Yau fourfolds constructed in the same way (as complete intersections in products of projective spaces) have also been recently classified \cite{Gray:2013mja}, giving rise to 921,947 configuration matrices.

There is another easy generalization to defining equations inside projective space and that is to change the scaling rule that led to compactness. For example rather than $\left( z_0,z_1,z_2\right) \sim \lambda \left( z_0,z_1,z_2\right)$ for $\mathbb{P}^2$, we could instead impose
\begin{equation}
\left( z_0,z_1,z_2\right) \sim  \left(\lambda z_0,\lambda^2 z_1,\lambda^3 z_2\right)~.
\end{equation}
This called a "weighted projective space", in this case it would be denoted $\mathbb{P}_{123}$ where the subscript indicates the weights. As in the cases described above, once again the Calabi-Yau condition ($c_1=0$) links the weights of the projective coordinates with the degree of the polynomials: $\sum \textit{weights} = \textit{deg. of polynomials}$ \cite{Candelas:1989hd}. For example, $\mathbb{P}_{123}[6]$ is a Calabi-Yau $1$-fold. These ideas can be generalized still further with interesting $\mathbb{C}^*$-scaling relations on coordinates in an ambient space known as a ``toric variety". Examples include those we've seen so far as well as bundles over projective spaces. Simple scaling rules like those above and the ability in some cases to encode geometric data in the combinatorics of a polytope make this class of geometries linked to the subject of convex geometry and possible to analyze in a very systematic way. We'll not go into the details here, but many useful and readable introductions exist (see \cite{Bouchard:2007ik} for a brief introduction and \cite{Cox} for a more in depth treatment), but we encourage the reader to explore further as toric spaces provide a useful testing ground for a wide range of physical and mathematical problems.

It is also useful to realize that these toric constructions are intrinsically linked to string compactifications through their realization in Gauged Linear Sigma models (GLSMs) \cite{Witten:1993yc}, whose CY target spaces are toric complete intersections. The majority of examples in the literature to date have focused on Abelian GLSMs, but recent examples of non-Abelian GLSMs have also given rise to novel constructions of CY manifolds (See for example \cite{Caldararu:2017usq}).

With this quick survey of algebraic constructions in hand, we are now in a position to ask an important question about these Calabi-Yau backgrounds -- \emph{how many Calabi-Yau manifolds are there?} To begin to answer it, we need a few topological tools that can be used to distinguish complex manifolds. It's clear from the above constructions that there may be more than one algebraic way of describing the same space, and what we're concerned with is the fundamental properties of the geometry itself, not a particular description. For example, we'll see in a moment that the following two configuration matrices describing Calabi-Yau one folds are in fact the same manifold, the complex elliptic curve (which is the same as the two torus, $T^2$, as a real manifold):
\begin{eqnarray}
\mathbb{P}^2[3],\\
\mathbb{P}^{123}[6]
\end{eqnarray}
So, it is clear that a systematic study of string compactifications will require us to be able to characterize and count distinct Calabi-Yau geometries. The multiplicities of CY $n$-folds (where $n$ is the complex dimension) starts out promisingly constrained, but rapidly grows:

\begin{itemize}
\item \textbf{n=1:} Only one manifold, the elliptic curve (complex curve of genus, $g=1$, same topology as $T^2$).
 \item \textbf{n=2} Once again, only one manifold, the so-called $K3$ surface. This is the unique non-trivial CY $2$-fold (excluding the trivial case of $T^2 \times T^2$). Examples of $K3$ surfaces include $\mathbb{P}^3[4]$ if we wish to build it via an algebraic description as described above or it could be constructed as an elliptic fibration over $\mathbb{P}^1$.
\item \textbf{n=3}  Here's where the simplicity breaks down. There are on order of a billion known examples of CY 3-folds. Constructions began with the 7890 CICY 3-folds \cite{Candelas:1987kf} described above and using toric methods Kreuzer and Skarke \cite{KS} constructed about half a billion toric ambient spaces in which CY hypersurfaces can be defined (i.e. anticanonical hypersurfaces).
 
 \item ${\bf n \geq 4}$ Once again it is not known how CY $n$-folds there are with $n>3$, however it is known that this is a vast set. There are $921,497$ CICY 4-folds compared to 7,890 CICY 3-folds and no systematic classification of CY 4-folds constructed as toric hypersurfaces has been attempted\footnote{At the moment, it may be beyond the reach of current computing power to attempt such a classification.} (though some partial datasets have been generated \cite{Klemm:1996ts}).

 \begin{figure}[h]
\centering
\includegraphics[width=0.7\textwidth]{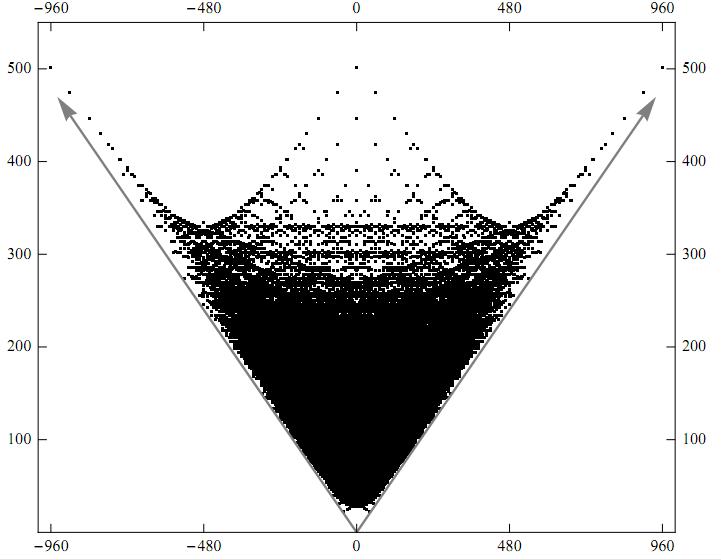}
\caption{\emph{The plot of Hodge numbers arising from the Kreuzer-Skarke dataset of Calabi-Yau threefolds defined as anti-canonical hypersurfaces in toric varieties \cite{KS}. The horizontal axis is $(h^{1,1}-h^{2,1})$ while the vertical axis is $(h^{1,1}+h^{2,1})$. Each point corresponds to a Calabi-Yau 3-fold with that (partial) topology, though many distinct manifolds may share those Hodge numbers. }}\label{Kreuzer_Skarke}
\end{figure}

\end{itemize}

To begin to understand this counting, we must now more seriously look at when two Calabi-Yau manifolds can be the same or different. For this we need the notion of topology, and deformation invariant numbers that can be used to characterize a manifold independent of a choice of metric. In general, the higher the complex dimension of a manifold, the more topological invariants that are needed to characterize it. The three quantities that will play a central role in this discussion are Chern classes, Hodge numbers (or more generally, bundle valued cohomology) and intersection numbers. These are all standard tools and we refer the reader to Appendix \ref{appendix} for a reminder of their definitions.\\

\newpage

\fbox{\begin{minipage}[c]{1\textwidth}
\textbf{How to tell if two Calabi-Yau manifolds are the same?} \\
 Topological invariants that can be used to distinguish Calabi-Yau n-folds: 
 \begin{itemize}
 \item \textbf{n=1} (Torus/Elliptic Curve)
 The genus of complex curves is the only topological invariant. 
 \begin{equation}
 g=1 \quad \Leftrightarrow c_1=0.
 \end{equation}
 \item {\bf n=2} ($K3$ surfaces) 
 \begin{equation}
 c_1=0,\quad c_2=24,\quad h^{1,1}=20.
 \end{equation}
 \item \textbf{n=3} (CY 3-fold) Wall's theorem \cite{wall} states that if for two Calabi-Yau manifolds the following collection of numbers are different then they are \emph{not diffeomorphic}: \\
\begin{equation}
(h^{1,1}(X),~h^{1,2}(X), ~c_2(X), ~D_a \cdot D_b \cdot D_c)
\end{equation}
\noindent where the last array of numbers are the triple intersection numbers of divisors in $X$.\\
The converse of the above statement is not true for complex manifolds, but it is true for real manifolds.
\item $\textbf{n=4}$ (CY 4-folds) Here the topological invariants include the following
\begin{eqnarray}
& c_2,\quad c_3,\quad c_4,\\
& h^{1,1},\quad h^{1,2},\quad h^{1,3},\quad h^{2,2}
\end{eqnarray}
\noindent with one linear relation between the Hodge numbers \cite{Klemm:1996ts}:
\begin{equation}
h^{2,2}(X)=2(22+2h^{1,1}(X)+2h^{3,1}(X)-h^{2,1}(X))
\end{equation}
A statement like Wall's theorem is expected to hold here (including appropriate intersection numbers), but is not yet proven. Since we have described the physical interpretation of the Hodge numbers of CY 3-folds in Section \ref{cy_overview}, it is worth a brief aside here to mention that within compactifications of M/F-theory, the Hodge numbers of a CY $4$-fold once again have a geometric/physical interpretation:
\begin{eqnarray}
 H^1(TX^{\vee}) &=&  H^{1,1}(X) \rightarrow \textit{Kahler moduli (Volumes)}\\
 H^1(TX) &=& H^{1,3}(X) \rightarrow \textit{Complex structure moduli}\\
 H^2(\wedge TX^{\vee}) &=& H^{2,2}(X) \rightarrow \textit{Relatd to G-flux in M/F-theory}
\end{eqnarray}
\item \textbf{n=5} (CY 5-folds) These are just beginning to be of interest in the literature (see \cite{Schafer-Nameki:2016cfr,Apruzzi:2016iac}). In principle they have six non-vanishing Hodge numbers $h^{1,1}(X),h^{2,1}(X), h^{1,3}(X), h^{2,2}(X),h^{1,4}(X),h^{2,3}(X)$, however like CY $4$-folds these obey one additional relationship \cite{Haupt:2008nu}:
\begin{equation}
11h^{1,1}(X)-10h^{2,1}(X)- h^{2,2}(X)+h^{2,3}(X)+10h^{1,3}(X)-11h^{1,4}(X)=0
\end{equation}
Here $h^{1,1}(X)$ and $h^{4,1}(X)$ are the K\"ahler and complex structure moduli respectively, while the rest parameterize various form fields similar to the $4$-fold case. \end{itemize}
\end{minipage}}\\

Let us know return to CY $3$-folds, our primary focus in this lecture series and what is being shown in Figure \ref{Kreuzer_Skarke}. It is clear from Wall's theorem that each datapoint in Kreuzer-Skarke Hodge number plot is definitely a different manifold, but because this does not include their Chern classes or triple intersection numbers there can be (and are!) many, distinct manifolds represented by the same point. In fact, the Kreuzer-Skarke dataset contains much more than half a billion manifolds, but the exact number is not known since the triple intersections (and full toric triangulations) have not been completed beyond low Hodge number examples (see \cite{Altman:2014bfa} for some recent efforts in this regard).

Inspection of the Kreuzer-Skarke dataset has led to many remarkable observations and questions. Let's explore a few briefly here:
\begin{enumerate}
\item \textbf{Mirror Symmetry} \cite{mirror_book}
The mirror symmetric nature of the plot in Figure \ref{Kreuzer_Skarke} makes visually obvious a very deep mathematical point (the starting point of homological mirror symmetry \cite{Kontsevich}). For CY threefolds, manifolds appear to come in pairs. For each $3$-fold $X$ with Hodge numbers $(h^{1,1},h^{2,1})$ there exists another manifold, $Y$, with that pair reversed \cite{Candelas:1990rm}:
\begin{eqnarray}
h^{1,1}(X) &=& h^{2,1}(Y) \\
h^{2,1}(X) &=& h^{1,1}(Y)
\end{eqnarray}
At the level of the string worldsheet this a fundamental symmetry of the sigma model. This observation has led to a tremendous number of advances in the physics of string compactions since many physical quantities that are difficult to compute in one Calabi-Yau geometry have a geometric analog in the mirror manifold which frequently makes the computation much simpler to do. 

Although CY mirror symmetry was discovered in CY threefolds, it has an analog for CY n-folds. There as, realized by the Hodge numbers, mirror symmetry implies $H^{p,q}(X)=H^{n-p,q}(Y)$.
\item \textbf{Finiteness?} One obvious question that we are immediately drawn to is that of finiteness. This has two layers: First, are there a finite number of Calabi-Yau threefolds? And second, are there are a finite number of possible values for the topological invariants (for example, Hodge numbers) of such threefolds? The answer to these questions is not yet known. It is clear that even if there are a finite number of Calabi-Yau threefolds, the total number of such manifolds is surely vast! Recall that the half a billion manifolds constructed by Kreuzer and Skarke in \cite{KS} were only \emph{hypersurfaces} in toric varieties. To put this in context, there are only 5 manifolds defined via a single polynomial (i.e. hypersurfaces) in the set of 7890 CICY threefolds. Interestingly, these 5 are given by configuration matrices that are the transpose of those associated to the cyclic CICYs described above:
\begin{equation}
\mathbb{P}^4[5] ~~ ,~~\begin{array}{c}
\mathbb{P}^1\\
\mathbb{P}^3
\end{array} \left[  \begin{array}{c}
2 \\
 4
\end{array} \right]~~ ,~~\begin{array}{c}
\mathbb{P}^2\\
\mathbb{P}^2
\end{array} \left[  \begin{array}{c}
3 \\
 3
\end{array} \right]
~~ ,~~\begin{array}{c}
\mathbb{P}^1\\
\mathbb{P}^1 \\
\mathbb{P}^2
\end{array} \left[  \begin{array}{c}
2 \\
 2 \\
 3
\end{array} \right]~~ ,~~\begin{array}{c}
\mathbb{P}^1\\
\mathbb{P}^1 \\
\mathbb{P}^1 \\
\mathbb{P}^1
\end{array} \left[  \begin{array}{c}
2 \\
 2 \\
 2 \\
 2
\end{array} \right]
\end{equation}
 That's 5 out of 7890. Now within the toric dataset, the 5 have grown to half a billion. How many more can we expect for the \emph{complete intersections (i.e. multiple polynomial defining relations)} in toric varieties? We don't know because they have yet to be classified (and it's doubtful that current computing tools are up to the task). Moreover, this is just one particular algebraic construction, by no means the most general way of building CY manifolds. So it's clear that the open questions of finitness of the whole set are difficult ones. But what about topology of CY threefolds? By inspection of the plot in Figure \ref{Kreuzer_Skarke} it can be observed that there is a substantial range in possible Hodge numbers:
 
Smallest topology: $(h^{11},h^{21})=(1,1)$\\
Largest topology: $(h^{11},h^{21})=(11,491)$

Perhaps more strikingly, in over 20 years of progress building CY manifolds and exploring their properties, the dataset of known Hodge numbers has stayed within the pattern laid out in Figure \ref{Kreuzer_Skarke}. That is, the known dots that might be added to that diagram have only been making it denser, but the plot of possible Hodge numbers hasn't got bigger. This includes the newest systematic construction of CY manifolds, known as gCICYs \cite{Anderson:2015iia}) which we'll discuss more in Section \ref{gcicys}. Take a look at \cite{skarke,benjamin} for useful websites summarizing known results about CY topology.
\item {\bf Geometry of the Hodge Number Plot?} It is clear that there is some structure to the shape\footnote{P. Candelas has compared it to the silhouette of a Texas longhorn cow.}  of Figure \ref{Kreuzer_Skarke}. Closer inspection of the diagram shows that the chain of points which generate the top outline of the diagram are repeated throughout the whole diagram in many different strands. This substructure has been the subject of some interest \cite{Taylor:2012dr,Candelas:2012uu}.  We'll be come back to this in a moment in Section \ref{fibrations} where we explore fibrations within CY threefolds. For the moment though, this leads us to our next question.
\item {\bf What geometric sub-structures are possible within this dataset?} For example
\begin{itemize}
\item Fibrations (which play a crucial role in F-theory compactifications \cite{Vafa:1996xn} and are an essential part of all string dualities (see e.g. \cite{Anderson:2016cdu}).
\item Blow-ups, contractible cycles (for example "Swiss cheese" \cite{Balasubramanian:2005zx,Cicoli:2008va} geometries in type IIB etc). The existence of certain divisors and/or curves can play a key role in the phenomenology of the resulting string compactification. \end{itemize} 
We'll see more of this substructure in detail soon in Section \ref{fibrations}. This is also related to our next question as the existence of sub-structure and especially contractible cycles (i.e. subspaces) can be used to connect distinct Calabi-Yau manifolds.
\item {\bf Is it possible to move \textit{dynamically} between the manifolds} in Figure \ref{Kreuzer_Skarke} (and hence their associated 4-dimensional theories)? The notion of a \emph{geometric transition} between two Calabi-Yau manifolds\footnote{Or indeed between Calabi-Yau manifolds and more general $SU(3)$ structure manifolds \cite{s3yau}.} is a variation of the geometry through some singular limit in which the topology of the manifold can change once the manifold is made smooth again\footnote{Note that variations of the K\"ahler and complex structure moduli of a Calabi-Yau threefold \emph{cannot} change topology if they are varied in such a way that the manifold stays smooth.}. There are really two related questions here. The first is a) Is it possible \emph{geometrically} to change one Calabi-Yau manifold into another and b) Is such a topology changing transition realizable in the associated field theory? (This is what's referred to as a 'dynamical' geometric transition in the wording of the question above). The latter is an attractive possibility as it would make the question of "which compactification background leads to which field theory?" somehow easier in that they would all be part of the same broader theory and vacuum space. In some string compactifications (notably type II theories) this type of \emph{dynamical} geometry changing transition is indeed possible and the rich interplay of geometry and field theory has been made very clear. For example, so-called conifold transitions \cite{Candelas:1989ug} were beautifully realized by the understanding that geometric singularities can lead to (a finite number of) extra light states in type IIA theories in \cite{Strominger:1995cz} (and that perhaps the existence of these extra light states can drive the effective physics \cite{Kofman:2004yc}). However, for many types of string compactifications (notably in the ${\cal N}=1$ effective heterotic string theory which we have used to illustrate techniques here), while it is understood that geometries can be linked mathematically, a full physical description of a topology changing transitions remains elusive.

Two of the most important types of topology changing transitions in string compactifications are
\begin{itemize}
\item Flop transitions (in which the Kahler moduli spaces of two different Calabi-Yau manifolds can be connected on a boundary in which volumes of sub-cycles collapse to zero size).
\item Conifold transitions (which involve the collapsing of an $S^2$ which can be replaced by an $S^3$, or vice versa) within the (real) $6$-dimensional Calabi-Yau manifold.
\end{itemize}
Between them, these two types of transitions connect just about all the Calabi-Yau threefolds that have been built with simple geometric constructions. 
\end{enumerate}
\subsection{Reid's Fantasy} 
Focusing for a moment on the mathematical process of geometric transitions, it has been conjectured by the mathematician Miles Reid \cite{reid} that \emph{all Calabi-Yau 3-folds are connected by mechanisms above like those above}. If true this would be a remarkable statement about that this broad class of string backgrounds are part of the same vacuum space. This remains however, a conjecture in general\footnote{Playing it safe, "fantasy" is used here by mathematicians to refer to a statement that is even more conjectural than a conjecture.}. There is evidence though for Reid's Fantasy within many datasets of CY manifolds. For example all CICY 3- and 4-folds in products of projective spaces can be connected via conifold transitions. 
 
\subsubsection{Conifold transitions}\label{conifold_sec}
Lets look at geometric transitions in a little more detail. The example we'll give here is a classic one \cite{Candelas:1989ug} but serves to simply illustrate the idea that distinct Calabi-Yau manifolds (with distinct topology) can be connected at singular points in their moduli space. 

Let's begin by considering the following two manifolds:

The first is the very simplest algebraic CY manifold, the quintic hypersurface with Hodge numbers $(h^{1,1},h^{2,1})=(1,101)$:

\begin{eqnarray}\label{smooth_quintic}
&\mathbb{P}^4[5]^{1,101},\nonumber\\
P_5 &= a_0x_0^5+a_1x_0^4x_1+\dots=0.
\end{eqnarray}
and the following co-dimension 2 manifold with Hodge numbers $(h^{1,1},h^{2,1})=(2,86)$:

 \begin{eqnarray}\label{Conifold of P45}
&\begin{array}{c}
\mathbb{P}_x^1 \\
\mathbb{P}_y^4
\end{array} \left[  \begin{array}{cc}
1 & 1 \\
1 & 4
\end{array} \right]^{2,86}\nonumber \\
P&= x_0 l_1(y)+x_1 l_2(y)=0,\nonumber\\
Q &= x_0 g_1(y)+x_1 g_2(y)=0.
\end{eqnarray}
wgere $l_i(y)$ and $g_j(y)$ are linear and quartic polynomials, respectively, in the coordinates of $\mathbb{P}^4$. These two polynomial equations can be written in matrix form as
\begin{equation}\label{matrix_form}
\left(  \begin{array}{cc}
l_1 & l_2 \\
g_1 & g_2
\end{array} \right)\left(\begin{array}{c}
x_0 \\
x_1
\end{array}\right)=0
\end{equation}
Because the solution $x_0=x_1=0$ is excluded in projective space, (\ref{Conifold of P45}) has a solution if and only if the determinant of the matrix in (\ref{matrix_form}) vanishes, which leads to the following degree 5 polynomial in $\mathbb{P}^4$:
\begin{equation}\label{singular_quintic}
l_1(y)g_2(y)-l_2(y)g_1(y)=0~.
\end{equation}
So, the two CICY manifolds given above both correspond to a quintic in $\mathbb{P}^4$. It is natural to ask...are they the same manifold? The answer is in fact no, because the quintic defined by (\ref{singular_quintic}) is singular. The singularities are at the common zeros of the $l$'s and $g$'s, which are 16 points.

\begin{figure}[h]
\centering
\includegraphics[width=0.8\textwidth]{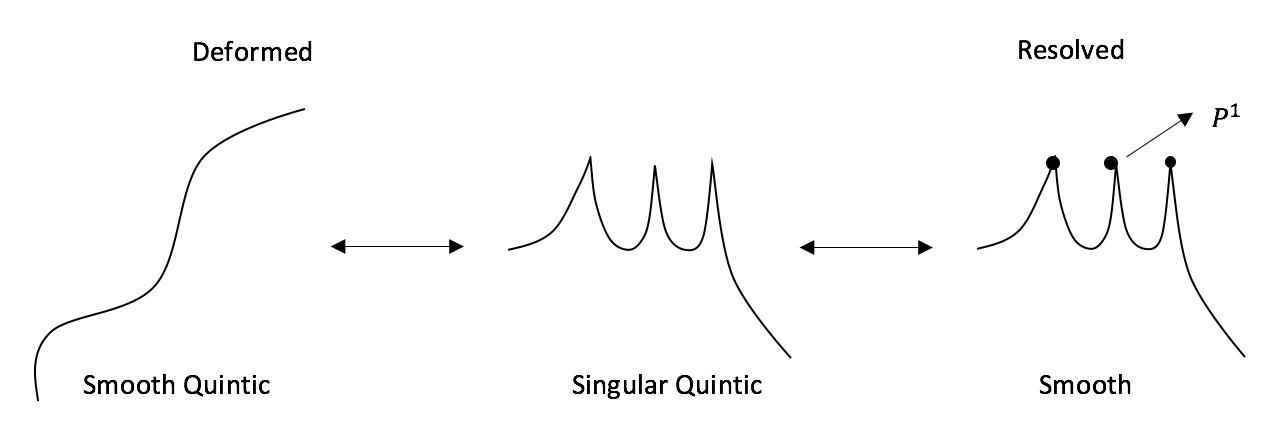}
\caption{\emph{An illustration of the classic conifold of the quintic \cite{Candelas:1989ug}.  On the left the manifold defined by a smooth quintic is shown, corresponding to the deformation side of the conifold. On the right is the resolution side of the conifold, corresponding to the resolved, co-dimension two smooth 3-fold.}}\label{Conifold}
\end{figure}
These manifolds are not the same, but share a common, singular locus in their moduli space. To move between these distinct geometries is an example of a so-called "conifold transition" \cite{Candelas:1989ug}. To reach the intermediate singular locus from the quintic side we must tune complex structure moduli (i.e. coefficients of the defining polynomials) until the quintic takes the form given in (\ref{singular_quintic}). Geometrically this tuning of complex structure corresponds to collapsing 3-cycles (take a look at \cite{Candelas:1990pi} to understand why $h^{2,1}$ moduli correspond to $S^3$'s in CY 3-folds). Once at that point, the singularities can be resolved by replacing each of the 16 singular points with an $S^2$ (in fact, those $S^2$'s are visible as the $\mathbb{P}^1$ which forms the difference between (\ref{singular_quintic}) and (\ref{Conifold of P45}). This changes the Euler number of the 3-fold by 32 in going from (\ref{smooth_quintic}) to (\ref{Conifold of P45}) (each added $\mathbb{P}^1$ contributes $2$ to the Euler number of the 3-fold). Phrased differently, while at the singular locus there are two branches of geometry (and indeed vacuum space of the associated string theory) meeting. One can move to the smooth quintic (called the "deformation" side of the conifold) or to the smooth co-dimension 2 CICY in (\ref{Conifold of P45}) (called the "resolution" side) by choosing which type of cycle to make large (i.e. in field theory, which type of field gets a vev). This simple example is the prototype for many similar transitions of this type, which can be linked together into chains and very possibly may connect all CY 3-folds.

\subsection{Fibrations in Calabi-Yau manifolds}\label{fibrations}
As we described previously, a natural structure to investigate within a Calabi-Yau manifold is the existence of a fibration. These are quite constrained within Calabi-Yau manifolds. Let $X$ be a Calabi-Yau threefold, then if $X$ can be written as a fibration $\pi: X \to B$ (with fiber $F$), it must fall into one of the following three types\footnote{Within a CY manifold, any fiber must also satisfy $c_1=0$, i.e. must also be a Calabi-Yau manifold by adjunction.} (see e.g. \cite{diverio}):
\begin{enumerate}
\item $B$ a surface and $F$ a genus one curve.
\item $B=\mathbb{P}^1$ and $F$ a K3 surface.
\item $B=\mathbb{P}^1$ and $F$ an abelian surface.
\end{enumerate}
Since fibrations play a key role in string dualities and force many simplifying features on the geometry and associated field theory, it is natural to ask how commonly they arise? What can be said about the existence of fibrations within known Calabi-Yau datasets?

To begin, it is important to note that the existence of a fibration 
\begin{equation}
\pi : X \rightarrow B
\end{equation}
is deformation invariant (i.e. topological) for Calabi-Yau n-folds with $n \ge 3$ (see \cite{Kollar})\footnote{Note that this is not the case for $n=2$ where deformations of a $K3$ surface can add or remove a fiber.}. So, for CY threefolds it actually makes sense to try to count what fraction of manifolds admit a fibration. Moreover, there is one crucial mathematical result, which makes elliptically fibered manifolds important within the set of all CY 3-folds: \\

\emph{The number of genus one\footnote{A bit of useful terminology here: a torus fibration that admits a section -- i.e. a map $s: B \to X$ -- is called ``elliptically fibered", while those more general geometries without a section are called ``genus one" (or $T^2$) fibrations.} fibered Calabi-Yau 3-folds is finite (see \cite{gross,grassi}).} \\

Although the proof of this statement is not constructive, however there have been serious efforts made to systematically enumerate this finite set of geometries \cite{Morrison:2012np}. The idea of the proof is related to the minimal model program \cite{hacon} and consists of a minimal set of base surfaces, $B_2$, consisting of the Enriques surface, $\mathbb{P}^2$, and the Hirzebruch surfaces ($\mathbb{F}_n$ with $0 \geq n \geq 12$) which can be blown up to produce new bases within the finite set (and this is used in \cite{Morrison:2012np} to form chains of non-Higgsable clusters). Similar constructions/finiteness results are also currently underway for CY 4-folds (see e.g. the recent work of di Cerbo and Svaldi \cite{dicerbo_svaldi}.)

It is now natural to ask whether or not this finiteness result can be related to the question of finiteness of all Calabi-Yau 3-folds? We would like to know
\begin{itemize}
\item What fraction of CY 3-folds are genus one fibrations?
\item Can general CY 3-folds be connected (via geometric transitions) to X with genus one fibrations?
\end{itemize}
To begin to answer these question, we'll first take a look at known datasets. We'll begin with asking about how many how many CY manifolds are fibered? In principle we could ask about any fiber type in the list above, but here I'll focus on $T^2$-fibers.

The results are as follows:
\begin{itemize}
\item For CICY threefolds more than $99\%$ admit at least one genus one fibration \cite{Anderson:2017aux,Anderson:2016cdu}

\item For CICY fourfolds, more than $99\%$ admit a genus one fibration \cite{Gray:2014fla}

\item For toric hypersurfaces, this has been studied somewhat less systematically than in the CICYs above, but preliminary studies indicate that at least $90\%$ are genus one fibered \cite{Braun:2011ux,rohsiepe,Johnson:2014xpa,Johnson:2016qar,Candelas:2012uu}.

\end{itemize}
For these known datasets of CY manifolds, the results are pretty striking! We can ask though whether this is a `lampost effect' in that perhaps all the CY manifolds we know how to build thus far (simple algebraic constructions in Fano ambient spaces) may exhibit special features? One bit of evidence that this is not the case, and that perhaps fibration structures really are ubiquitous comes from the newest, systematic construction of CY manifolds -- the so-called gCICYs \cite{Anderson:2015iia}. We will explore this construction in more detail in Section \ref{gcicys}, but for now we will note that although gCICYs have not yet been fully classified (only a small proof-of-principle dataset was constructed in \cite{Anderson:2015iia}), so far, they all admit fibrations with the same frequency as the CICY dataset.

Genus one fibrations in CY threefolds appear to not only be very common, but they also play an important role in bounding the known datasets. As we mentioned in the previous section, the Kreuzer-Skarke Hodge number plot in Figure \ref{Kreuzer_Skarke} has a curious shape to its upper boundary. In fact, every one of those manifolds is genus one fibered and are all connected via geometric transitions (similar to those discussed in Section \ref{conifold_sec}) \cite{Taylor:2012dr}. Indeed, the extremal known Hodge numbers (the extremal dots in Figure \ref{Kreuzer_Skarke}) of $(h^{1,1},h^{2,1})=(11,492)$ (or the mirror) are not only the largest Hodge numbers appearing in the set of toric hypersurfaces, they have also been proved to be the maximal Hodge numbers for \emph{any elliptically fibered CY threefold}, using arguments from $6$-dimensional compactifications of F-theory \cite{Taylor:2012dr}.

It appears then that nearly all known CY manifolds are genus one fibered. But what about those that aren't? The $1\%$ of manifolds that do not admit any fibration are not scattered randomly throughout the datasets, they all share the common feature of very small $h^{1,1}$ (for example, the quintic clearly does not admit an elliptic fibration). Within the Kreuzer-Skarke plot in Figure \ref{Kreuzer_Skarke}, the non-fibered geometries all lie close to the tip of the diagram. Within the CICY threefolds, every manifold with $h^{1,1}>4$ admits a genus one fibration. More generally, it appears within known datasets that every Calabi-Yau threefold with $h^{1,1} > 19$ admits a genus one fibration. This leads us to conjecture an intriguing possibility: \\

\noindent {\bf Conjecture\footnote{Actually this is more at the "fantasy" level again.}: }Every Calabi-Yau threefold (regardless of construction) with $h^{1,1}>19$ is genus one fibered. \\

If true, this would have some profound consequences for the finiteness of all CY threefolds. Specifically, it would immediately tell us that the Hodge numbers $(h^{1,1},h^{2,1})=(11,492)$ are maximal for all CY threefolds (see \cite{Keller:2012mr} for some hints in this direction from string sigma models). Moreover, this could be used to bound all CY threefolds systematically if Reid's fantasy is correct and all non-fibered manifolds could be connected to fibered ones via geometric transitions. This is all very conjectural still, but it makes it clear that the problem of classifying CY backgrounds of string theory is not a hopeless one. The sub-structure of these manifolds provides a rich playground in which we can attempt to study and bound them. For now, let's look at this fibration structure in a little more detail.

Hunting for fibrations in a CY manifold can involve several different approaches and some of the searches cited above are exhaustive, while some have only been looking for so-called "Obvious" fibrations, which are manifest within a given algebraic construction of the manifold. We'll begin by making this latter class of fibrations explicit.
 
\subsubsection{Obvious genus one fibrations}
 Within a CICY threefold, one way to spot a fibration structure is via the configuration matrix itself \cite{Anderson:2017aux}. For example, if any CICY of the general form (\ref{cy-config}) can be arranged in the following block diagonal form
 \begin{equation}\label{T2fibration}
 { X}= \left[\begin{array}{c|c:c}  {\mathcal{A}}_1 & 0 & T^2 \\ \hdashline
{\cal A}_2 & {\textit{Base}} & {\textit{Twist}} \end{array}\right] .
\end{equation}
 then it is $T^2$ (i.e. genus one) fibered. In the above, ${\cal A}_1,{\cal A}_2$ indicate ambient spaces (for example products of projective spaces), while the blocks denoted ``$T^2$, base, and twist" indicate multi-degrees of polynomials. Those of the upper right block satisfy the conditions to give a Calabi-Yau one-fold, while the base gives a (non-CY) twofold base to the genus one fibration (and the remaining 'twist' multi-degrees indicate how the $T^2$ fiber is non-trivially twisted over the base).

For example, this threefold,
 \begin{equation}\label{obv_fib}
X=\left[\begin{array}{c|c:c}  \mathbb{P}^2 & 0 & 3 \\ \hdashline
\mathbb{P}^3 & 2 & 2 \end{array}\right] .
\end{equation}
defines a fibration $\pi: X \to B$ where the fiber is described by $\mathbb{P}^2[3]$ and the base is $B=\mathbb{P}^3[2]$. To see why this is a fibration, consider solving the first hypersurface equation (i.e. the first column in (\ref{obv_fib})) to obtain the surface $\mathbb{P}^3[2]$. Pick any point on this surface and plug it into the second equation (i.e second column of (\ref{obv_fib})). What remains is a cubic polynomial in $\mathbb{P}^2$ with specific coefficients (i.e. complex structure of the genus one curve). If we vary the point on $\mathbb{P}^3[2]$ that we pick, we end up with a different defining polynomial for $\mathbb{P}^2[3]$. This is exactly what it means to be a fibration! For each point in the base ($\mathbb{P}^2$) there is a genus one curve $\mathbb{P}^2$ over it. As we move over the base, the part of the defining polynomials (labeled by the "twist" degrees above) changes the form of the $T^2$ over that point. This type of ``obvious" fibration structure is quite simple to search for algorithmically and such scans have been done for the CICY 3-folds and 4-folds \cite{Gray:2014fla,Anderson:2016cdu,Anderson:2017aux}. Moreover, there are analogous ways to spot projection maps in toric data \cite{Braun:2011ux} that make fibration scans in toric hypersurfaces possible, as well as the toric ``top" construction (see e.g. \cite{Candelas:2012uu}). It is these approaches that have made possible the statements given above that most known CY manifolds admit fibrations.

It turns out however, that there is another observation waiting to be made using the same observations above. Which is that almost all known Calabi-Yau n-folds are not just fibered, they admit \emph{more than one distinct fibration!}

Let's look at an example of a random CICY configuration matrix, corresponding to a threefold that admits a fibration, $\pi: X \to B$ where $B$ is a multi-degree $\{1,1,1\}$ hypersurface in $\mathbb{P}^1 \times \mathbb{P}^1 \times \mathbb{P}^1$ (also known as the third del Pezzo surface, $dP_3$):
\begin{equation}\label{multieg}
X=\left[\begin{array}{c|c:cc}  
\mathbb{P}^1 & 0 & 1 & 1 \\ 
\mathbb{P}^2 & 0 & 1 & 2\\ \hdashline
\mathbb{P}^1 & 1 & 0 & 1 \\
\mathbb{P}^1 & 1 & 0 & 1 \\
\mathbb{P}^1 & 1 & 0 & 1
 \end{array}\right] .
\end{equation}
Here base and fiber is defined respectively as,
\begin{equation}
B = dP_3=\left[ \begin{array}{c|c}
\mathbb{P}^1 & 1 \\
\mathbb{P}^1 & 1 \\
\mathbb{P}^1 & 1
\end{array}\right], \quad T^2=\left[ \begin{array}{c|cc}
\mathbb{P}^1 & 1 & 1\\
\mathbb{P}^2 & 1 & 2
\end{array} \right]
\end{equation}
But this is not all that is possible! An interesting thing about CICY configuration matrices of the form (\ref{cy-config}) is that it is possible to permute the rows and columns without changing the geometry (permuting columns corresponds to switching the order in which the complete intersection is written and permuting columns changes the order of the ambient $\mathbb{P}^n$ factors, neither of which changes what is meant by the complete intersection geometry). If we perform such row/column swaps on (\ref{multieg}) it is possible to re-write this manifold as
\begin{equation}
X=\left[\begin{array}{c|ccc} 
\mathbb{P}^1 & 1 & 0 & 1 \\ 
\mathbb{P}^1 & 0 & 1 & 1 \\ 
\mathbb{P}^2 & 1 & 0 & 2 \\ \hdashline
\mathbb{P}^1 & 0 & 1 & 1 \\
\mathbb{P}^1 & 0 & 1 & 1
 \end{array}\right] .
\end{equation}
Here base and fiber is defined respectively as,
\begin{equation}
B = \mathbb{F}_0=\mathbb{P}^1 \times \mathbb{P}^1 , \quad T^2=\left[ \begin{array}{c|ccc}
\mathbb{P}^1 & 1 & 0 & 1\\
\mathbb{P}^1 & 0 & 1 & 1\\
\mathbb{P}^2 & 1 & 0 & 2\\
\end{array} \right]~.
\end{equation}
Note that here the block denoted as ``base" in (\ref{T2fibration}) is trivial. The base is just $\mathbb{P}^1 \times \mathbb{P}^1$ and picking a point in those two ambient factors leaves us with the equation of a torus, given by the co-dimension three configuration matrix given above. This new description -- as a fibration $\tilde{\pi}: X \to \mathbb{F}_0$ -- brings with it detailed information about the manifold, its substructure, how it can go singular, etc. The more descriptions as fibrations like this that are available, the more information there is to be had! We'll return to the consequences of these multiple fibration structures for string dualities in a moment.

First, it should also be realized that one manifold can admit multiple fibrations of other types as well. For example, CY 4-folds can have multiple descriptions as $K3$-fibered or $CY$ 3-fold fibered manifolds. In fact, all possible ``nestings" of fibrations are possible and generically one manifold can admit many. That is, we may see fibers nested as
\begin{equation}
T^2\subset K3 \subset CY3 \subset CY4
\end{equation}
To illustrate this, consider the following two descriptions of the same threefold:
\begin{equation}
\left[\begin{array}{c|c:cc} 
\mathbb{P}^1 & 0 & 1 & 1 \\ 
\mathbb{P}^2 & 0 & 1 & 2 \\ \hdashline
\mathbb{P}^1 & 2 & 0 & 0 \\ 
\mathbb{P}^1 & 1 & 1 & 0 \\ \hdashline
\mathbb{P}^1 & 1 & 0 & 1
 \end{array}\right] \quad \textit{Vs.} \quad \left[ \begin{array}{c|ccc}
\mathbb{P}^1 & 1 & 1 & 0 \\ 
\mathbb{P}^1 & 1 & 0 & 1 \\ 
\mathbb{P}^2 & 1 & 2 & 0 \\ \hdashline
\mathbb{P}^1 & 0 & 0 & 2 \\ \hdashline
\mathbb{P}^1 & 0 & 1 & 1
\end{array}  \right].
\end{equation}
There is only one CY threefold here, but it can be written as several different fibrations. In the left configuration matrix, the dotted lines highlight a $K3$ fibration which itself is a torus fibration over $\mathbb{P}^1$,
\begin{equation}
T^2 = \left[ \begin{array}{c|cc}
\mathbb{P}^1 & 1 & 1 \\
\mathbb{P}^2 & 1 & 2\\
\end{array}\right] \subset \quad K3=\left[ \begin{array}{c|c:cc}
\mathbb{P}^1 & 0 & 1 & 1 \\ 
\mathbb{P}^2 & 0 & 1 & 2 \\ \hdashline
\mathbb{P}^1 & 2 & 0 & 0 \\ 
\mathbb{P}^1 & 1 & 1 & 0
\end{array} \right].
\end{equation} 
In contrast, in the configuration matrix given on the right, we have highlighted a different fibration structure. This one corresponds to a different elliptic fibration of the manifold, within the same $K3$ fibration.
\begin{equation}
T^2 = \left[ \begin{array}{c|ccc}
\mathbb{P}^1 & 1 & 1 & 0 \\ 
\mathbb{P}^1 & 1 & 0 & 1 \\ 
\mathbb{P}^2 & 1 & 2 & 0 
\end{array}\right] \subset \quad K3 = \left[ \begin{array}{c|ccc}
\mathbb{P}^1 & 1 & 1 & 0 \\ 
\mathbb{P}^1 & 1 & 0 & 1 \\ 
\mathbb{P}^2 & 1 & 2 & 0  \\ \hdashline
\mathbb{P}^1 & 0 & 0 & 2
\end{array} \right].
\end{equation}
That is, the one $K3$-fibration of the threefold contains two distinct elliptic fibrations. This type of nested fibration structure plays an important role in string dualities. See \cite{Anderson:2016cdu} for a recent summary.

 \subsection{Enumerating Fibrations}
 The fibration structures described in the previous sections were all of a form that was easy to see within a given algebraic description of the manifold. If we want to simply survey whether known CY manifolds admit fibrations, this is already quite a lot of information. However, we may still ask -- even if a manifold does not obviously admit an elliptic fibration in a given description, is it possible that it still fibered? After all, one CY manifold can have more than one algebraic description. In addition, is there a way to count absolutely how many inequivalent  such fibrations for one CY n-fold? The answer to the latter question is yes, thanks to results in the mathematics literature due to Oguiso and Wilson for CY 3-folds \cite{oguiso,wilson} (and conjectured to hold to for $n>3$ by Koll\'ar \cite{Kollar:2012pv}). The conjecture can be stated as follows: \\
 
 \noindent Conjecture \cite{Kollar:2012pv}: \emph{Let $X$ be a Calabi-Yau n-fold. Then $X$ is genus one fibered if and only if there exists a divisor $D$ such that $D \cdot C \geq 0$ for every algebraic curve $C \subset X$, $D^{dim(X)}=0$ and $D^{dim(X)-1} \neq 0$.} \\
 
\noindent Here $D^{dim(X)}$ denotes the $n$-intersection of $D$ with itself, etc. In \cite{oguiso,wilson} this conjecture was proven for CY 3-folds subject to the additional constraint that $D$ is effective or $D \cdot c_2(X) \neq 0$. Intuitively this criteria is characterizing the existence of a fibration by characterizing a particular divisor in the base manifold of that fibration. The role of the divisor $D$ above is that of a pull-back of an ample divisor in the base, $B$, (where the fibration is written $\pi: X \to B$). For more details, we refer the interested reader to \cite{Kollar:2012pv,Anderson:2017aux}.
 
 This tool allows us to now enumerate \emph{all genus one fibrations} of a CY n-fold, rather than just the obvious one. Such a systematic study was carried out for the CICY 3-folds in \cite{Anderson:2017aux}, where more than $377,559$ genus one fibrations were found for the $7,890$ manifolds. For the known datasets of CY 3-folds, it turns out not only is the generic manifold fibered, but that the average number of such fibrations is $\sim 10$ (and for CICY 4-folds that average goes up to $\sim 100$). For some remarkable CICY 3-folds, there are $\sim 10,000$ fibrations in one manifold. In fact, for one well-known CY 3-fold -- the so-called ``split bi-cubic" or ``Schoen manifold" with Hodge numbers $(h^{1,1},h^{2,1})=(19,19)$
 \begin{equation}
\left[ \begin{array}{c|cc}
\mathbb{P}^1 & 1 & 1\\
\mathbb{P}^2 & 3 & 0 \\
\mathbb{P}^2 & 0 & 3
\end{array} \right]
 \end{equation}
it can be demonstrated that there are an \emph{infinite} number of distinct $T^2$-fibrations \cite{oguiso_inf}. This is in contrast to the mere $4$ obvious genus one fibrations\footnote{Exercise for the reader: Can you find them?} visible from the description above.

This rich and plentiful array of fibration structures is remarkable! Mathematically, this sub-structure (i.e. dividing the manifold as fiber+base in a number of different ways) let us study complicated manifolds in some generality. For example, decomposing manifolds into fiber and base allows us to classify possible geometries \cite{gross} and derive patterns in their sub-structure, intersection numbers, etc. Physically, as we saw above they play a deep role in string dualities and in mapping out the redundancy in the string landscape. This has implications for the questions we raised in Section \ref{intro} about which effective theories can arise in string compactifications.

It should be noted though that perhaps the ubiquity of fibration structures is a feature of how we've learned how to build CY manifolds thus far. To shed a small amount of light on that question we will take a brief detour to explore the newest construction of CY manifolds (where once again, we will find an abundance of fibrations).

\subsection{``Generalized" Complete Intersection Calabi-Yau manifolds (gCICYs)}\label{gcicys}
For the most part, the constructions of CY manifolds we've described above have all had a simple form -- namely the manifolds were described by a complete intersection of simple polynomial equations in a Fano ambient space. It turns out that this type of algebraic construction can be easily made a little more general than one might expect. The result is a complete intersection in a non-Fano ambient space, which has been called a ``Generalized Complete Intersection" CY manifold \cite{Anderson:2015iia}.

To begin, let's look at a simple and important observation for the early days of string compactifications: \emph{algebraically realized CY manifolds can have so-called ``non-polynomial" deformations}. Recall that in Section \ref{alg_sec} (near equation (\ref{fermat})) we counted the deformations of the defining polynomial of the quintic and showed that (modulo coordinate redefinitions) it exactly matched the expected $101$ complex structure moduli of the 3-fold. This is not always the case however. Take for example, the following CICY realized by two polynomial equations
\begin{equation}
 \begin{array}{c}
\left[ \begin{array}{c|cc}
\mathbb{P}^1 & 0 & 2\\
\mathbb{P}^4 & 3 & 2
\end{array} \right] \\
h^{2,1}= 56\\
h^{1,1}= 2
\end{array}
\end{equation} 
If we perform the same counting here as we did for the quintic, we find only $51$ inequivalent polynomial deformations. So somehow there are $5$ missing \emph{non-polynomial} deformations in this case. The existence of these non-polynomial deformations was a bit of a mystery in the literature. It leads us to ask
 \begin{itemize}
 \item Is there a simple geometric origin of these non-polynomial deformations?
 \item Would it be possible to build a CY manifold with \emph{entirely} non-polynomial deformations?
 \end{itemize}
 It was these questions that were addressed in \cite{Anderson:2015iia} and lead to a new dataset of CY manifolds known as gCICYs. Here is an example of one such 3-fold:
\begin{equation}\label{eg_gcicy}
X_3= \left[ \begin{array}{c|cc|cc}
\mathbb{P}^1 & 1 & 1 & -1 & 1 \\
\mathbb{P}^1 & 1 & 1 & 1 & -1 \\
\mathbb{P}^5 & 3 & 1 & 1 & 1
\end{array} \right].
\end{equation} 
At first glance this looks very strange indeed. Since the rows of a configuration matrix of the form (\ref{cy-config}) are supposed to denote the polynomial multi-degrees of the defining equations in the given ambient space, what does this notation even mean?
 
Let's begin on the column which seems to be describing an equation with multi-degree $(-1,1,1)$. There is clearly no such polynomial on $\mathbb{P}^1 \times \mathbb{P}^1 \times \mathbb{P}^5$. However, there could be one on this manifold
 \begin{equation}\label{m_def}
 \mathcal{M} = \left[ \begin{array}{c|cc}
 \mathbb{P}^1 & 1 & 1 \\
\mathbb{P}^1 & 1 & 1  \\
\mathbb{P}^5 & 3 & 1 
\end{array}  \right]
 \end{equation}
 Mathematically, this is the statement that  the cone of \emph{effective divisors} (i.e. divisors that can be written algebraically) is larger on $\mathcal{M}$ than on $\mathbb{P}^1 \times \mathbb{P}^1 \times \mathbb{P}^5$. In the coordinates of $\mathcal{M}$ a polynomial defining equation could look ordinary. However, that does not guarantee that it's so simple in the coordinates of $\mathbb{P}^1 \times \mathbb{P}^1 \times \mathbb{P}^5$.

In these simple projective space coordinates, what does an equation of degree $(-1,1,1)$ look like? It is the combination of a divisors of zeros and a divisor of poles. That is, it takes the form of a numerator over a denominator (where the negative degrees specify the denominator)
\begin{equation}
\frac{\mathcal{N}^{(0,1,1)}(y,z)}{\mathcal{D}^{(1,0,0)}(x)}
\end{equation}
This is a meromorphic, not holomorphic function in the coordinates of the ambient product of projective spaces. However once we have defined the positive degree equations to make the manifold $\mathcal{M}$ in (\ref{m_def}) above, not all meromorphic functions are created equal.

To begin exploring the possibilities, let's write the second defining equation of $\mathcal{M}$ explicitly as
\begin{equation}
x_0 P_1^{(1,1)}(y,z)+x_1 P_2^{(1,1)}(y,z)=0.
\end{equation}
Now, observe that when $x_0 = 0$ this equation enforces the fact that $P_2^{(1,1)}(y,z)=0$ (since $x_0=x_1=0$ is not allowed in projective space). Therefore if we choose
\begin{equation}
\frac{\mathcal{N}^{(0,1,1)}(y,z)}{\mathcal{D}^{(1,0,0)}(x)}=\frac{P_2^{(1,1)}(y,z)}{x_0}
\end{equation}
this divisor has no poles! That is, it's a holomorphic function on $\mathcal{M}$ and the zeros of the denominator ``miss" the hypersurface. Using these techniques the two negative columns in (\ref{eg_gcicy}) sweep out good (and quite specific) algebraic equations on $\mathcal{M}$ and $X_3$ is a good ``algebraic" CY (but clearly the ``order" of defining equations matters).

In an initial proof of principle scan, about $6000$ manifolds of this type were constructed in \cite{Anderson:2015iia} (also further studied in \cite{Anderson:2015yzz,Garbagnati:2017rtb} and the approach generalized still further in the toric context in \cite{Berglund:2016yqo,Berglund:2016nvh}). In that initial study new topology (i.e. Hodge numbers, Chern classes and intersection numbers) were found. Moreover, once again, nearly all of them admit fibrations (elliptic, K3, etc) and exhibit  new base manifolds compared to the ordinary CICYs.

So far, no classification of such manifolds has been constructed and it is clear that this will already be a difficult task. Oddly, it is very easy to construct seemingly infinite families of gCICYs. For example,
\begin{equation}
\left[ \begin{array}{c|c:c}
\mathbb{P}^1 & 2+i & -i \\
\mathbb{P}^4 & 1 & 4
\end{array} \right]
\end{equation}  
for all $i \geq 0$ is a good gCICY. However, we have not proven so easily that the number of CY threefolds is infinite. Closer inspection reveals that this infinite set is actually all describing the same geometry \cite{Anderson:2015iia}. It is in intriguing open area of exploration to try to learn more about manifolds constructed in this way.

\section{Role of bundles in heterotic theories}\label{bundle_sec} 
In the previous sections we have focused on the diverse possible background manifolds that can arise in string compactifications. In particular, we began in Section \ref{het_sec} to explore solutions to the heterotic Strominger system and thus far, we have been ignoring one crucial aspect of the vacuum structure -- namely, the presence of gauge fields. From the supersymmetry variations in Section \ref{lagrangian_etc}, we must return to one final condition for a supersymmetric vacuum to the theory, namely (\ref{bundle_eom}). To leading order this is 
\begin{equation}\label{bundle_eom2}
\Gamma^{MN}F^A_{MN}=0
\end{equation}
where $F^{A}_{MN}$ is the field strength, $A$ is a gauge index and $MN$ are spacetime indices (in real coordinates). 

For string compactifications to lower dimensions, we will generically be interested in solutions with a non-trivial background vev on $X$ the compactification geometry. The presence of such a $\langle A \rangle \neq 0$ will break the 10-dimensional $E_8$ gauge symmetry down to a subgroup, $G \subset E_8$ where $G \times H \subset E_8$ and $H$ is the gauge symmetry associated to the gauge fields over the compact directions. In this section we will focus on gauge fields on Calabi-Yau manifolds, $X$.

Geometrically, the gauge fields over $X$ are described as a \emph{vector bundle}, $\pi: V \to X$. Similar to the notion of fibrations we discussed in Section \ref{manifold_building}, a fiber bundle locally takes the form of ``Fiber $\times$ Base" (although globally there is a non-trivial twisting over a compact base) and consists of a non-trivial projection from a total space to the base. A vector bundle is a fiber bundle whose fiber space is also a vector space. Unlike the case of fibrations discussed before, the fibers of bundles do not degenerate anywhere over the base. For our purposes the base will consist of a Calabi-Yau manifold, $X$, while the fiber vector space is associated to a representation of a Lie group (see Appendix \ref{appendix} for details).

Returning now to (\ref{bundle_eom2}) we can phrase this as a constraint over the \emph{background} gauge fields on $X$ and hence on the connection of the vector bundle, $V \to X$. Written in terms of complex coordinates on $X$, we have the following pair of equations:
\begin{equation}\label{HYM}
\delta \chi \Rightarrow \left\lbrace \begin{array}{c}
F_{ab}=F_{\bar{a}\bar{b}}=0 \\
g^{a\bar{b}}F_{a\bar{b}}=0
\end{array} \right\rbrace 
\end{equation}
These are known as the Hermitian Yang-Mills equations. It is important to note from the start that unfortunately, these are wickedly hard PDEs to solve! (Recall that we don't even know the Calabi-Yau metric, $g^{a\bar{b}}$, explicitly and now must use it in solving a non-linear PDE for the gauge field\footnote{See \cite{Douglas:2006hz,Anderson:2011ed,Anderson:2010ke} for some numeric approaches to this problem.}). Fortunately, the solutions of these equations (a problem in differential geometry) is linked to properties of the vector bundle $V \to X$ which can be phrased in terms of algebraic geometry. These properties correspond as follows:
\begin{eqnarray}
 F_{ab}=F_{\bar{a}\bar{b}}=0 &\Leftrightarrow &V \textit{is holomorphic}\label{hym1}\\
 g^{a\bar{b}}F_{a\bar{b}}=0 &\Leftrightarrow & V \textit{is slope (Mumford) poly-stable} \label{hym2}
 \end{eqnarray}
The first of these conditions is easy to understand conceptually -- a bundle is called "holomorphic" if its transition functions are holomorphic functions over the complex base manifold (entirely analogously to the condition on transition functions that a manifold be complex, see Appendix \ref{appendix}). The second condition -- that of bundle ``stability" -- is a little harder to describe intuitively. The definition of stability is linked to the notion of ``slope", a geometric quantity (quasi-topological, depending on the first Chern class of the sheaf and the K\"ahler moduli of $X$) defined as
\begin{equation}
\mu(V)=\frac{1}{rk(V)}\int_X c_1(V) \wedge \omega \wedge \omega
\end{equation}
where $\omega$ is the K\"ahler form on $X$. A bundle is stable if the slope associated to all sub-sheaves of a bundle is strictly less than that of the bundle, $V$:
\begin{equation}
\mu({\cal F}) < \mu(V) ~~\forall~{\cal F} \subset V
\end{equation}
Moreover, \emph{poly-stable} bundles are direct sums of stable bundles, all with the same slope: $V=\bigoplus_i V_i$ with $\mu(V_i)=\mu(V)$ $\forall i$. For the interested reader, the full definition of bundle stability and it's relationship to the Hermitian Yang-Mills equations is given in Appendix \ref{appendix}. For now, it is enough to note that unfortunately, we are frequently faced with "conservation of misery", in that much like solving (\ref{HYM}), proving that a bundle is stable is also a highly non-trivial task! (See \cite{Anderson:2008ex} for some analytic and \cite{Douglas:2006hz,Anderson:2011ed,Anderson:2010ke} numerical techniques that have been recently applied in string theory).

As was noted in Section \ref{het_sec}, the gauge fields in a heterotic theory must satisfy an anomaly cancellation condition (\ref{anom_canc_orig}):
 \begin{equation}\label{anomalycancellation}
 dH \sim tr(F\wedge F)-tr (R\wedge R)
 \end{equation}
An important consequence of this fact is that the existence of a vector bundle over a compactification manifold, $X$, is \emph{not optional} in a heterotic theory. There is no way to ``turn off" the bundle. Instead, we must choose non-trivial gauge field vevs over $X$ and guarantee that the background connection/bundle satisfies the Hermitian Yang-Mills equations. This can be difficult in general to do, though there is always one solution available. The so-called ``standard embedding" is to take the vector bundle to be the holomorphic tangent bundle to the Calabi-Yau manifold $X$ itself, i.e. $V=TX$. In this case the spin connection is set equal to the gauge connection and $H=0$ automatically. Although such solutions were the prototypical heterotic solutions \cite{Candelas:1985en}, they are far from the only possibilities. We will explore other possible bundles and bounds below.

The expression in (\ref{anomalycancellation}) can be integrated over 4-cycles in $X$ to provide a necessary condition on the second Chern characters of the bundle and the manifold (the Bianchi identity):
\begin{equation}\label{bianchi}
ch_2(V)+[C]=ch_2(TX)
\end{equation}
where $[C]$ is the class of an effective curve, $C$. This is added to the anomaly cancellation condition to include the possiblity of NS 5-branes in the heterotic theory (equivalently M5-branes in heterotic M-theory). Thus far we have not really touched on these non-perturbative contributions to the theory. For the present purposes, it is enough to note that through \emph{heterotic small instanton transitions}, 5-branes can be viewed simply as degenerations of the fibers of the vector bundle (i.e. the bundle degenerates into a sky-scraper sheaf). These ``small instantons" can be deformed back into smooth bundles or moved into the interval directions of heterotic M-theory in a process known as a small instanton transition. Such possibilities are intricate and important and though we will not explore them in detail here, we refer the reader to \cite{Ovrut:2000qi} for more details.

In the following section, we will investigate how the presence of $\left\langle A_{\mu}\right\rangle$ on $X$ changes and enriches the problem of classifying background geometry, determining the moduli of the system, and characterizing the effective physics.

\subsection{Heterotic Massless Spectra}
In addition to the dilaton and the geometric moduli associated to the Calabi-Yau manifold ($h^{1,1}(X)$ and $h^{2,1}(X)$), the $10$-dimensional $E_8$ gauge fields will give rise to sector of particles in the low-energy theory. Because of the quite restricted form of the 10-dimensional heterotic lagrangian (\ref{10D_lag}), the only origin of charged matter in the $4$-dimensional effective theory is the gauge field in 10-dimensions. That is, all $4$-dimensional matter must descend from the 248 dimensional adjoint representation of $E_8$. 

As described above, the low energy gauge group $G$ in the 4-dimensional theory is given by the commutant of the bundle structure group $H \subset E_8$. For example, for
$H={\rm SU}(3)$, ${\rm SU}(4)$, ${\rm SU}(5)$ this implies the standard
grand unified groups $G=E_6$, ${\rm SO}(10)$, ${\rm SU}(5)$, respectively in 4-dimensions.
In order to find the matter field representations, we have to
decompose the adjoint ${\bf 248}$ of $E_8$ under $G\times H$. In
general, this decomposition can be written as
\begin{equation}
{\bf 248}\rightarrow (\mbox{Ad}(G),{\bf 1})\oplus \bigoplus_{i}(R_{i},r_{i})  
\label{adjoint}
\end{equation}
where $\mbox{Ad}(G)$ denotes the adjoint representation of $G$ and
$\{(R_{i},r_{i})\}$ is a set of representations of $G\times H$.
The adjoint representation of $G$ of course corresponds to low-energy
gauge fields, while the low-energy matter fields transform in the representations
$R_i$ of $G$. Examples of this decomposition are given in Table \ref{spec}.

The decomposition above describes the type of matter that \emph{could} arise, not what necessarily \emph{does} arise in the string compactification. To get the true number of massless degrees of freedom in the theory, we must count bundle-valued harmonic forms. That is, the number of massless modes for a given representation is counted by the dimension of a certain bundle-valued cohomology group \cite{GSW2}. To get an intuitive feel for why this is the case, consider part of the equation of motion described above (\ref{HYM}) under fluctuations, $A \rightarrow A + \delta A$:
\begin{equation}
F_{\bar{a}\bar{b}}=0 \Rightarrow \overline{D}(\delta A)=0
\end{equation}
where $\overline{D}$ is the gauge covariant derivative. That is, all fluctuations of the connection must be $\overline{D}$ closed. Since exact contributions to such a fluctuation are pure gauge, it is clear that we need to count closed forms modulo exact forms, i.e. cohomology (see Appendix \ref{appendix}).

Now, we can make a little more precise how this leads to $4$-dimensional matter fields. Consider the following more explicit form for the fluctuation of the gauge connection over the compact internal space:
\begin{equation}
A_{\overline{a}}=A^{0}_{\overline{a}}+\delta C^{i}_{x}T^{x \alpha} \omega_{i \bar{a}\alpha}
\end{equation}
where $A^{0}_{\overline{a}}$ is the background choice of gauge field on the CY manifold, the coefficient of the fluctuation, $\delta C$ is a $4$-dimensional matter field, $T^{x \alpha}$ is a generator of $G \times H$, and $\omega_{i \overline{a}\alpha}$ is a basis of harmonic forms\footnote{Satisfying $d{\omega}= \star d{\omega}=0$.} on $X$. The index $x$ runs over the range of the representation of the group $G$ while $\alpha$ ranges over the dimension of the representation of $H$ and $i$ is a flavor index ($\overline{a}$ remains the spatial index on the CY manifold). 

To summarize all this, what we're observing is that the number of charged $4$-dimensional matter fields $\delta C$ are counted by the forms $\omega$ carrying indices of the internal CY manifold, $X$, and the gauge degrees over $X$, i.e. the $H$-bundle over $X$. Thus, the number of $4$-dimensional supermultiplets occurring in the low energy theory for
each representation $R_i$ is given by
$n_{R_{i}}=h^{1}(X,V_{r_{i}})$, where $R_{i}, r_{i}$ are defined by the decomposition~\eqref{adjoint}.
For $H={\rm SU}(n)$, the relevant representations $r_i$ can be obtained
by appropriate tensor and anti-symmetric products of the fundamental representation. The relevant cohomology groups and hence the number of low-energy representations can then be computed as
summarized in Table \ref{spec}.
\begin{table}[h]
\begin{center}
\begin{tabular}{|l|l|l|}
\hline
$G\times H$ & Breaking Pattern:  
${\bf 248}\rightarrow $ 
& Particle Spectrum\\  
\hline\hline
{\small $ E_{6}\times \rm{SU}(3)$} & {\small $({\bf
  78},{\bf 1})\oplus ({\bf 27},{\bf 3})\oplus 
(\overline{\bf 27},\overline{\bf 3})\oplus ({\bf 1},{\bf 8})$ }
&
$
\ba{rcl}
n_{27}&=&h^{1}(V_3)\\ 
n_{\overline{27}}&=&h^{1}(V^\vee_{\bar{3}})=h^{2}(V_3)\\
n_{1}&=&h^{1}(V_3\otimes V_{\bar{3}}^\vee)
\ea
$
\\  \hline
{\small $\rm{SO}(10)\times\rm{SU}(4)$} &{\small $({\bf
  45},{\bf 1})\oplus ({\bf 16},{\bf 4}) 
\oplus (\overline{\bf 16},\overline{\bf 4})\oplus ({\bf
  10},{\bf 6})\oplus ({\bf 1},{\bf 15})$ } 
&
$
\ba{rcl}
n_{16}&=&h^{1}(V_4)\\
n_{\overline{16}}&=&h^{1}(V_{\bar{4}}^\vee)=h^2(V_4)\\
n_{10}&=&h^{1}(\wedge ^{2}V_4)\\
n_{1}&=&h^{1}(V_{4}\otimes V_{\bar{4}}^\vee)
\ea
$
\\  \hline
{\small $\rm{SU}(5)\times\rm{SU}(5)$} &{\small $ ({\bf 24},{\bf 1})\oplus
({\bf 5},\overline{{\bf 10}})\oplus (\overline{\bf 5},{\bf 10})\oplus ({\bf 10},{\bf 5})\oplus 
(\overline{\bf 10},\overline{\bf 5})\oplus ({\bf 1},{\bf 24})
$}
&
$
\ba{rcl}
n_{10}&=&h^{1}(V_5)\\ 
n_{\overline{10}}&=&h^{1}(V_{\bar{5}}^\vee)=h^2(V_{5})\\ 
n_{5}&=&h^{1}(\wedge^{2}V_{5}^\vee)\\
n_{\overline{5}}&=&h^{1}(\wedge ^{2}V_{5})\\
n_{1}&=&h^{1}(V_{5}\otimes V_{\bar{5}}^\vee)
\ea
$
\\ \hline
\end{tabular}
\caption{{\em A vector bundle $V$ with structure group $H$ can break the $E_8$
  gauge group of the heterotic string into a $4$-dimensional GUT group $G$. The low-energy representation are found from the branching of the ${\bf 248}$ adjoint of $E_8$ under $G\times H$ and the low-energy massless spectrum is obtained by computing the indicated bundle cohomology groups.}}\label{spec}
\end{center}
\end{table}

As a consequence of this counting of zero modes by cohomology, it follows that the Atiyah-Singer index computes the chiral asymmetry in the 4-dimensional theory. When $c_1(TX)=c_1(V)=0$, the index of $V$ can be expressed as 
\begin{equation}
Ind(V)=\sum_{p=0}^3 (-1)^{p}\, h^{p}(X,V)=\frac{1}{2}\int_{X}c_3(V) \ ,
\label{index}
\end{equation}
where $c_3(V)$ is the third Chern class of $V$. For a stable $SU(n)$ bundle
we have $h^0(X,V)=h^3(X,V)=0$. As a result, comparison with Table \ref{spec} shows
that the index counts the chiral asymmetry, that is,
the difference of the number of generations and anti-generations. 

\subsection{The potential of the ${\cal N}=1$, 4-dimensional theory}
Finally, the structure of the potential of the $4$-dimensional, ${\cal N}=1$ theory is related to the geometry of the vector bundle and the form of the zero modes as counted above. The potential of the ${\cal N}=1$ supersymmetric theory is determined by the K\"ahler and superpotentials \cite{Wess:1992cp} as
\begin{equation}
V=e^{K}\left[K^{A{\bar B}}F_{A}{\bar F}_{\bar B}-3|W|^2 \right]
\end{equation}
where sub- and superscripts denote derivatives, $K$ is the Kahler potential, $W$ the superpotential (a holomorphic function of the fields), and the ``F-terms" are given by $F_{A}=W_{A}+K_{A}W$.

Historically, the form of the matter field K\"ahler potential has remained largely unknown in heterotic compactifications (as a complicated function of the CY metric and Hermitian Yang-Mills connect), while the superpotential could be more explicitly analyzed. However, some recent progress \cite{Blesneag:2018ygh,Blesneag:2015pvz,Candelas:2016usb} has been made in understanding the form the K\"ahler potential. More straightforwardly, the superpotential is closely related to the cohomological origin of the 4-dimensional fields and to deformation problems in the underlying geometry \cite{Berglund:1995yu,Anderson:2011ty,Anderson:2013qca}. For example, the tri-linear couplings of $4$-dimensional fields are related to Yoneda (cup) products in cohomology (i.e.  $\sim$ $H^1(X,V) \cup H^1(X,V)  \cup H^1(X,V)$):
\begin{equation}
\lambda_{ijk} \phi^{i}\phi^{j}\phi^{k}~~~\Rightarrow~~\lambda_{ijk}\sim \int_{X} {\omega^{i}}_{\bar{ a}}{\omega^{j}}_{\bar {b}}{\omega^{k}}_{\bar {c}}\Omega^{{\bar a}{\bar b}{\bar c}}
\end{equation}
where ${\omega^{i}}_{\bar{ a}}$ is a bundle valued one-form and $\Omega$ is the holomorphic $(0,3)$ on $X$. For example, in the case of a ${\bf 27}^3$ coupling in an $E_6$ theory, ${\omega^{i}}_{\bar{ a}} \in H^1(X, V_3)$.

In addition, non-perturbative contributions to the superpotential (for example due to worldsheet instantons or gaugino condensation) are also linked to geometry in that their support is over rational curves in $X$ which can be explicitly constructed and enumerated. See e.g. \cite{Buchbinder:2017azb,Anderson:2011cza} for recent examples.

In previous sections we have focused on constructions and classifications of the compactification manifolds. We will turn now to how such problems change in the presence of gauge fields (or fluxes etc) on $X$ and how they can impact the deformations and effective physics of the system.

\subsection{Classifying pairs $(X,V)$ for heterotic backgrounds}
It is an interesting observation that gauge fields and the manifolds that they exist over can constrain one another in interesting ways. For example, as we saw in Section \ref{manifold_building} there are precisely two line bundles that can be defined over the circle, $S^1$ -- the cylinder and the M\"obius strip. In the complex fiber case, it follows that there are only these two ways of defining a $U(1)$ gauge theory (whose fiber corresponds to the gauge degree of freedom, $\phi$ in the gauge transformation $A \to A + d\phi$) on a circle. Now, CY manifolds are far more complicated geometries than $S^1$, but it is clear that here too, the possible gauge theories that can be written down are highly constrained by the form of the compact geometry.

Let us brielfly review here the conditions that must be placed on heterotic bundles and what can be said about how many such objects there are. Clearly, the bundle on $X$ must be holomorphic and stable, as we saw in (\ref{hym1}) and (\ref{hym2}) which is equivalent to satisfying the Hermitian Yang-Mills equations. So it is natural to ask \\
 
 \textbf{Classification problem:} How many stable, holomorphic bundles can arise in a heterotic compactification on a Calabi-Yau manifold $X$? Can we bound the topology of the bundle? \\

Here by topology of the bundle we include the Chern classes $c_1(V)$, $c_2(V)$, $c_3(V)$. As we have seen above, each of these is related to a physical aspect of the theory. These include the existence of spinors, anomaly cancellation conditions, and the chiral index of the theory, respectively. These quantities also come with some bounds imposed by consistency of the theory 
 \begin{itemize}
 \item $c_1(V)=0$ since the structure group of $V$ must be a sub-group of $E_8$ and the slope, $\mu(V)$ must vanish from stability and (\ref{hym2}). Moreover, the condition that spinors can exist in the $4$-dimensional theory is controlled by the vanishing of the second Steifel-Whitney class, which here indicates that $c_1(V)=0$ (mod 2)\footnote{Note that in the case of a poly-stable bundle, $V_{total}=\oplus V_i$, the net $c_1(V_{total})=0$, while $c_1(V_i)$ may be non-vanishing, with $\mu(V_i)=0$.} \cite{GSW2}. 
 \item As described in (\ref{bianchi}) anomaly cancellation indicates that $c_2(V) \leq c_2(TX)$ (i.e. given a choice of $X$, the second Chern class of $V$ is bounded).
 \item As in (\ref{index}) the third Chern class is linked to the chiral asymmetry: $\frac{1}{2} c_3(V)=Ind(V)$ $\Rightarrow$ number of generations $-$ number of anti-generations. Interestingly, this number is known to be linked to the other two Chern classes for stable bundles: \\
 
 \textbf{Theorem} \cite{maruyama,langer}:For a stable bundle, $V$, if $c_1(V)$ and $c_2(V)$ are fixed, then there exists a finite number of possible values for $c_3$. \\
 
 This indicates that given a choice of $X$ there are in principle a finite number of topologies available for $V$.
  \end{itemize}

Much as in Section \ref{manifold_building}, where we described methods to build manifolds, techniques for building vector bundles likewise build from simple pieces up to more complicated forms. The basic building block is the Abelian bundle, i.e. a line bundle, which on a Calabi-Yau manifold is classified by its first Chern class (written $L={\cal O}(a_i D_i)$ where $i=1,\ldots h^{1,1}(X)$, $D_i$ is a basis of divisors on $X$ and $c_1(L)=a_i \omega^i$ where $\omega^i$ are a basis of K\"ahler forms on $X$). Line bundles can be combined to form non-Abelian bundles\footnote{In addition, much recent progress in heterotic constructions has been obtained through using fully Abelian bundles to ``probe" non-Abelian moduli spaces \cite{Anderson:2011ns,Anderson:2012yf,Anderson:2013xka,Anderson:2014hia}.}  through a variety of means including monad constructions \cite{Okonek} and extensions \cite{Hartshorne}. We'll touch on these briefly below. For now though, we turn to a more careful look at the moduli in the heterotic theory.

 \subsection{Manifolds, Bundles and Moduli Problems}
 In this section we will take another look at the moduli -- both geometric and physical -- that arise in heterotic compactifications. It is clear that the relevant geometric fluctuations are those of the metric on $X$ and the connection on $V$. However, the way that these two types of deformations can influence each other can be subtle and important. We'll explore this in detail.
 
 We touched on the fluctuations of the CY metric in Section \ref{cy_overview} where we summarized the standard result that in the absence of gauge fields or flux, the moduli of a CY manifold are parameterized by $h^{1,1}(X)$ K\"ahler moduli and $h^{2,1}(X)$ complex structure moduli. Now we can ask, what about the moduli associated to the vector bundle?
 
The moduli of the bundle are derived by considering fluctuations, $A\rightarrow A+\delta A$, which preserve the defining equations (where of course $A$ is valued in adjoint representation of $G$). Beginning with (\ref{HYM}) we can consider such a fluctuation of the connection while holding the metric on $X$ fixed. To first order this leads to
 \begin{eqnarray}
 F_{ab}=F_{\bar{a}\bar{b}}=0, \qquad \Rightarrow \bar{D}_{\bar{a}}(\delta A_{\bar b})=0
 \end{eqnarray} 
That is, the bundle fluctuations which preserve the holomorphic structure of $V$ are closed under the bundle-valued covariant derivative operator, $\bar{D}$. Moreover, taking into account the fact that exact fluctuations are pure gauge in this case, we see that
 \begin{equation}
\textit{Bundle Moduli} \Rightarrow H^1(X,End_0 (V))
\end{equation}
Note that this is a space of 1-forms since $\delta A$ has one spacetime index over the CY manifold and is adjoint valued in $G$ (this is denoted in representations of the principle bundle by taking the traceless endomorphism bundle $End_0(V)$ associated to $V$ in the fundamental representation). See also Appendix \ref{appendix} for more on cohomology groups and what it means for forms to be closed modulo exact pieces.

This observation led to the following naive count (appearing frequently in the literature since the 1980s) of massless singlets in a heterotic theory on a CY background:
\begin{equation}\label{naive_count}
h^{1,1}(X)+h^{2,1}(X)+h^1(X,End_0(V)).
\end{equation} 
But the astute reader will observe that in the calculation of the bundle moduli we held the metric fixed and likewise when the metric fluctuations were considered, we did not include the gauge fields. This is not the way fluctuations really work (in reality we must vary all fields at once). So we must ask, do the geometric fluctuations really separate like in (\ref{naive_count})? Can the bundle and manifold constrain each other? (Note that we could ask similar questions for other structures similar to bundles, such as m-forms on X, gerbes etc.).
 
 Returning to the equations arising from supersymmetry variations, it is clear that $F_{ab}=F_{\bar{a}\bar{b}}=0$ and $g^{a\bar{b}}F_{a\bar{b}}=0$ depend on the complex structure and K\"ahler structure of $X$, as well as the fluctuations of the gauge connection. There are two questions we'd like to address here: \\
 
 \textbf{Question 1:} What is the total heterotic moduli space? (Where the metric and connection both vary at once).  \\
 
 \textbf{Question 2:} If we start with a solution and vary in a direction so that the HYM equations are \emph{not} satisfied, \emph{what happens in $4$-dimensional effective theory?} \\
 
We will begin with Question 2. To analyze the form of the 4-dimensional theory, it is helpful to consider a part of the 10-dimensional action as was done in \cite{Anderson:2011ty}:

\begin{equation}
S_{partial} \sim \int_{M_{10}} \sqrt{-g} \left( tr(F)^2-tr (R)^2 + \ldots \right)
\end{equation} 
Now, these terms can be re-written using information for the 10-dimensional Bianchi identity:
\begin{equation}
dH\sim tr(F\wedge F)-tr(R\wedge R)\quad \Rightarrow \quad \int_X \omega \wedge \left(tr(F\wedge F)-tr(R\wedge R) \right)=0.
\end{equation}
where $\omega$ is the K\"ahler form on $X$. Next, using the fact that $X$ is Ricci flat and K\"ahler to the first order leads to
 \begin{equation}
 \Rightarrow \int_{X} \sqrt{-g} \left( tr(F)^2-tr(R)^2 +2tr(F_{a\bar{b}}g^{a\bar{b}})^2-4tr(F_{ab}F_{\bar{a}\bar{b}}g^{a\bar{a}}g^{b\bar{b}})\right)=0
 \end{equation}
Substituting this into $S_{partial}$ yields:
\begin{equation}
S_{partial} \sim \int_{X} \sqrt{-g}\left\lbrace (-\frac{1}{2}tr(F_{a\bar{b}}g^{a\bar{b}})^2+tr(F_{ab}F_{\bar{a}\bar{b}}g^{a\bar{a}}g^{b\bar{b}}) \right\rbrace
\end{equation}  
It is now possible to observe that the above terms in the 10-dimensional theory contain no 4-dimensional derivatives and thus, must be a contribution to the potential of the 4-dimensional theory. Moreover, they contribute positive semi-definite terms to that potential. If the Hermitian Yang-Mills equations are satisfied, this part of the 4-dimensional potential is zero. If however, fields are fluctuated so that geometrically the Hermitian Yang-Mills equations are \emph{not} satisfied, then this potential becomes non-trivial. More precisely the failure/success of a bundle to be slope-stable leads to a $4$-dimensional D-term \cite{Sharpe:1998zu,Anderson:2009sw,Anderson:2009nt} while the property of bundle holomorphy corresponds to F-terms \cite{Anderson:2010mh,Anderson:2011ty}.

To return to the questions phrased above, can we derive this potential explicitly? What are the true moduli of a heterotic theory? Let's start with the easy equations: 
\[F_{ab}=F_{\bar{a}\bar{b}}=0. \]
 What happens if we start with a solution, and then vary the complex structure of X? Must the bundle stay holomorphic? It turns out the answer is no! To see this, it is helpful to re-write the equations above in terms of real coordinates (so we can observe the change in complex structure explicitly). To this end, we introduce the projection operators
 
\begin{equation}\label{Projections}
P_i^j = (\mathbb{I}_i^j +i J_i^j),\qquad \bar{P}_i^j=(\mathbb{I}_i^j -i J_i^j)
\end{equation}
with $J^2=-\mathbb{I}$, the complex structure tensor and $i,j$ real coordinates on $X$. Then the Hermitian Yang-Mills equations in (\ref{HYM}) in terms of real coordinates will be equivalent to the following trio of conditions:
\begin{eqnarray}\label{realcrdHYM}
g^{ij}P_i^k \bar{P}_j^l F_{kl} &=& 0, \nonumber \\
P_i^k P_j^l F_{kl} &=& 0, \\
\bar{P}_i^k \bar{P}_j^l F_{kl} &=& 0. \nonumber
\end{eqnarray}
With this in hand, it is now possible to consider the simultaneous perturbations of the complex structure on $X$ and the connection on $V$:
\begin{equation}
J=J^{(0)}+\delta J,\qquad A=A^{(0)}+\delta A.
\end{equation}
It was shown in \cite{Anderson:2010mh,Anderson:2011ty} that in terms of complex coordinates, $\delta J_a^{\bar{b}}= -i \bar{v}_{I a}^{\bar{b}} \delta z^I$, with $v_I$ are a basis of tangent bundle valued, harmonic 1-forms and $\delta z^I$ a fluctuation of complex structure, are the only non-zero components of $\delta J$.

Plugging these fluctuations in to (\ref{realcrdHYM}), we obtain the following constraint (to first order in the fluctuated fields):
\begin{equation}\label{atiyah_eqn}
\delta z^I v_{I[\bar{a}}^c F^{(0)}_{c|\bar{b}]}+2\bar{D}^{(0)}_{[\bar{a}}\delta A_{\bar{b}]}=0
\end{equation}
This is the equation that takes the place of a constraint on fluctuations of either the complex structure or the connection alone. It is easy to understand the origin of this expression. The left term corresponds to the rotation of the $(1,1)$ component of the field strength to the $(0,2)$ component due to the change in complex structure. The right term, corresponds to the new contribution in $(0,2)$ component from a non-closed fluctuation $\delta A$. It is clear that everything we counted as a good bundle modulus before still satisfies this equation since $\delta A \in H^1(X, End_0(V))$ is closed under $\bar{D}_{\bar a}$. But what about the complex structure fluctuations? Not every $\delta z^I$ can be balanced by a $\delta A$! 

It is clear that we need to characterize solutions to (\ref{atiyah_eqn}) in order to determine the number of true moduli in the theory. Fortunately, this geometric question was addressed in the 1950s by Atiyah \cite{atiyah}. To understand his result, it is useful to review three objects in deformation theory:
\begin{enumerate}
\item ${\bf Def(X)}$: 
Deformations of $X$ as a complex manifold. Infinitesimal (i.e. 1st order) deformations correspond to $H^1(TX) = H^{2,1}(X)$ in the case of a CY 3-fold $\Rightarrow$ Complex structure moduli.
\item ${\bf Def(V)}$:
Deformations of $V \rightarrow X$ (i.e. changes in the connection, $\delta A$) for a fixed complex structure of $X$. Infinitesimal deformations correspond to $H^!(X,End_0(V))$ $\Rightarrow$ Bundle moduli.
\item ${\bf Def(X,V)}$:
Simultaneous holomorphic deformations of $X$ and $V$. Here the infinitesimal (1st order) simultaneous deformations of the manifold/bundle pair correspond to $H^1(X,Q)$. Where the bundle $Q$ is defined by
\begin{equation}\label{Q}
0\rightarrow End_0(V) \rightarrow Q \xrightarrow{\pi} TX \rightarrow 0.
\end{equation}
This sequence is known as the \emph{Atiyah Sequence} (see the definition of a short exact sequence near (\ref{ses}) below for a summary of the notation). It was demonstrated in \cite{Anderson:2010mh,Anderson:2011cza,Anderson:2011ty} that $H^1(X,Q)$ are the actual complex moduli of a heterotic theory.
\end{enumerate} 
 
 \begin{centering}
\fbox{\begin{minipage}{40em}

 \textbf{Short Exact Sequences (SES):} 
 
\begin{equation}\label{ses}
0 \xrightarrow{f_1} A \xrightarrow{f_2} B \xrightarrow{f_3} C \xrightarrow{f_4} 0
\end{equation} 
 
Let $A, B, C$ be bundles and each map satisfy Ker$(f_{n+1})$ $=$ Im$(f_n)$ for all n. It follows that $A \subset B$, and $C= B/A$.  

Therefore $B$ is ``almost" $A \oplus C$, but not quite. $A$ and $C$ are non-trivially ``glued ". The connection on $B$ takes the form 
$\left( \begin{array}{cc}
\mathcal{A}_A & Ext^1 \\
0 & \mathcal{A}_C
\end{array} \right)$. Where the gluing data corresponds to $Ext^1(C,A)=H^1(X, A\otimes C^{\vee})$ (when $A$ and $C$ are bundles). Short exact sequences are a crucial tool in algebraic geometry.

 \end{minipage}}
 \end{centering}
 
From the long exact sequence in cohomology associated to (\ref{Q}) it can be derived that 
\begin{equation}\label{atiyah_result}
H^1(X,Q)=H^1(End_0(V)) \oplus ker (\alpha),
\end{equation} 
where $\alpha=F_{a{\bar b}}^{0}$ is the so-called ``Atiyah Class" (and the $\{1,1\}$ component of the field strength in background).
 \begin{equation}\label{AtiyahClass}
 \alpha : H^1(TX) \rightarrow H^2(End_0 (V)), \qquad \alpha=[F^{1,1}_{0}] \in H^1 (TX^{\vee} \otimes End_0(V)).
 \end{equation}
 An element of $H^1(X,Q)$ is exactly the solution to the fluctuation problem laid out above. Note that all the bundle moduli, $H^1(X, End_0(V))$ are a subset of $H^1(X,Q)$ automatically. However, \emph{not all complex structure moduli of $X$ will be included} (and in general the sum in (\ref{naive_count}) is an over counting!). An element of $ker(\alpha) \subset H^1(TX)$ is a variation of the complex structure who's image is zero under the schematic mapping $\delta z F_{a{\bar b}}$. But care needs to be taken since the target space is a cohomology group, ``zero" in cohomology simply means exact. That is, schematically
$ \delta z F_{a{\bar b}} \sim \bar{D}(\delta A)$.

 This is exactly the same as the condition we derived in (\ref{atiyah_eqn}) above! Thus, the true complex moduli of a heterotic theory are the naive bundle moduli in $H^1(X, End_0(V))$ and a subset of the complex structure moduli of the base manifold, $X$. The intuitive notion is that the presence of the gauge fields can ``freeze" complex structure of the base manifold. Geometrically, the complex structure variations described by the simultaneous deformation space $(X,V)$ can also be derived by considering the ordinary complex structure of the projectivization of the total space of the bundle $\mathcal{P}=\mathbb{P}(\pi:V \to X)$. In this case, an analysis using a Leray spectral sequence shows that $H^1({\cal P}, T{\mathcal P})$ reduces to $H^1(X, Q)$ on the base \cite{donaldson_def}.

With these observations in hand, it is important to now ask how many complex structure moduli of $X$ can be ``frozen" by a bundle? How can we compute the dimension $h^1(X,Q)$? This is actually difficult to do in general because the Atiyah class is nothing less than $[F^{1,1}_0]$, the background gauge field strength. As described above, very few explicit examples are known of HYM connections or field strengths. This has led to very few explicit calculations of an Atiyah deformation space in the literature. As we will see below though, some progress can be made.

Let us begin with simple constructions of bundles and ask how many complex structure moduli of $X$ they can constrain:
 \begin{itemize}
\item Abelian gauge fields: $L \rightarrow X$ . Here, $H^1(End_0(V)) = 0$, so no restriction on moduli.
\item Simplest non-abelian example is an extension (i.e. non-trivial ``gluing") of two line bundles to give a non-abelian bundle. For example an $SU(2)$ bundle can be constructed via the following short exact sequence:
\begin{equation}\label{su2_ext}
0 \rightarrow L \rightarrow V \rightarrow L^{\vee} \rightarrow 0, \qquad c_1(V)=0.
\end{equation}
This is non-trivial only when $Ext^1(L^{\vee},L)=H^1(X,L^{\otimes 2}) \ne 0$ (see the form of the connection below (\ref{ses}). When the extension class in $Ext^1(L^{\vee},L)$ is chosen to be zero, this corresponds to the ``split" bundle $L \oplus L^{\vee})$.
\end{itemize}
As it turns out this simple construction of $SU(2)$ bundles\footnote{This class of bundles can be used to make manifest the 4-dimensional D-terms related to slope stability \cite{Sharpe:1998zu,Anderson:2009sw,Anderson:2009nt}.} has been shown \cite{Anderson:2010mh,Anderson:2010ty,Anderson:2011cza,Anderson:2011ty,Anderson:2013qca} to fix nearly all the complex structure moduli of $X$! The idea is simple to state -- this construction of bundles manifestly depends on the complex structure of the base manifold. Above, the extension class $H^1(X,L^{\otimes 2})$ can actually ``jump'' in dimension with changes in the complex structure of $X$! The ``jumping'' here has the form that $H^1(X,L^{\otimes 2})=0$ generically in complex structure moduli, but can become non-zero for certain special, higher co-dimensional loci. As a result, it makes sense that this bundle can ``obstruct'' its base manifold $X$, since the ``building blocks'' of $V$ may only exist for certain complex structure of $X$ (here $Ext^1(L^{\vee},L)$). See \cite{Anderson:2011ty} for examples of this sort and generalizations.

If moduli are fixed because of gauge fields on $X$, how can this appear in the heterotic effective theory? In addition to the $10$-dimensional picture given in (\ref{atiyah_result}) above, this moduli fixing can also be understood from the point of view of the 4-dimensional heterotic theory directly. For heterotic compactifications on an $SU(3)$ structure manifold, the superpotential can be described via a Gukov-Vafa-Witten (GKV) form \cite{Gukov:1999ya}
\begin{equation}
W=\int \Omega \wedge H
\end{equation}
with $H=dB-\frac{3}{\sqrt{2}}\alpha ' (\omega^{YM}_3-\omega^{L}_3)$, $\omega_3 \sim tr(F \wedge A -\frac{1}{3} A\wedge A\wedge A)$.
and in the case of heterotic bundles in \cite{Anderson:2011ty} it was shown that Atiyah-type obstructions as shown above will give rise to F-terms from the GKV superpotential.
 
 This is especially easy to see schematically in the case of an $SU(2)$ bundle defined by extension in (\ref{su2_ext}). There, we observed that if the extension class $H^1(X, L^{\otimes 2})$ jumps, the bundle can constrain moduli of $X$. For CY manifolds such jumping must occur in a way that preserves the index given in (\ref{index}). That is, two types of line bundle cohomology must jump in dimension simultaneously. These give rise to 4-dimensional charged matter fields that we will label as
 \begin{equation}
 C_+ \in H^1(L^{\otimes 2}),\qquad C_- \in H^1(L^{\vee \otimes 2}).
 \end{equation}
 This jumping phenomenon can be realized simply in a superpotential of the form
\begin{equation}
W= \lambda_{ia}(z) C_+^iC_-^a + \Gamma_{ijab}C_+^iC_+^jC_-^aC_-^b+\dots  
\end{equation} 
Note that for generic values of the complex structure, the holomorphic function $ \lambda_{ia}(z)\neq 0$. As a result, the fields $C_+$ and $C_-$ are generically massive. However, on a special locus in complex structure moduli space, $ \lambda_{ia}(z)=0$ and the fields become massless. This corresponds the jump in dimension of the line bundle cohomology group described above. Once these fields are massless, they can be given a vev -- a process that geometrically corresponds to defining the non-Abelian $SU(2)$ bundle as above (see \cite{Anderson:2009nt} for more details).
If we choose $\left\langle C_+ \right\rangle$ to define the extension class given above, then the geometric process of constraining the complex structure moduli can be realized in the 4-dimensional theory as the simple F-term obstruction
 \begin{equation}
 \frac{\partial W}{\partial C_- ^a} = \lambda_{ia}(z) \left\langle C_+^i  \right\rangle
 \end{equation}
 near the locus in moduli space where $\lambda(z)=0$, in fluctuation $\delta z^I_{\perp}$ becomes massive (where $\delta z_\perp$ are the complex structure moduli directions away from the locus where $\lambda(z)=0$.
 
 The above arguments are just a quick sketch of careful correspondences that can be laid out between geometry and the underlying effective theory. The overall message that we hope to illustrate with this brief overview is that geometry and field theory agree! Any geometric obstruction has a consequence in field theory and it is crucial to consider complete geometric objects in understanding the effective low energy physics. In addition, choosing the geometry carefully (i.e. a selection of pairs $(X,V)$) can dramatically reduce the number of moduli in the theory (which may be of use for certain phenomenological applications). Above we have illustrated the effect of gauge fields $A_{\mu}$ on $X$, but similar tools could be employed to study non-trivial flux backgrounds in string compactifications (i.e. $\langle H_{\mu\nu\rho} \rangle \neq 0$).  In the heterotic literature these tools have been employed to study geometry and field theory in the Strominger system with non-Kahler geometry \cite{Anderson:2014xha,delaOssa:2014cia,delaOssa:2015maa}. Recent progress has also been made on Type IIB compactifications using the same approach \cite{Gray:2018kss}.
 
 \section{Conclusions}
 This has been a quick, curated overview of some results and open questions in string compactifications. The reader we had in mind was a non-expert who hoped to get a better feel for this subject and \emph{why} these questions are interesting and important (rather than an in-depth or comprehensive technical review). We hope that we will have enticed such a reader to explore further and there are many exciting directions that we have not had a chance to cover here. In particular, progress on non-K\"ahler compactifications and M-theory compactifications on $G_2$ manifolds is rapidly developing. We encourage the reader to continue to explore these and many other interesting open questions raised in these notes. Good luck!
 
 \section*{Acknowledgements}
 We thank the organizers of \emph{TASI 2017 -- ``Physics at the Fundamental Frontier"} for their hospitality and the opportunity to be a part of the excellent summer school in which these lectures were given. L.A. would also like to thank S.J. Lee and J. Gray for useful discussions.  The work of L.A. is supported in part by NSF grant PHY-1720321 and is part of the working group activities of the the 4-VA initiative ``A Synthesis of Two Approaches to String Phenomenology".

 \appendix
 \section{Appendix}\label{appendix}

In order to be self contained, we briefly review some of the main mathematical objects that are frequently used in string theory compactification, for more detailed information refer to \cite{GSW2,Huybrechts,Nakahara:1990th,Hubsch:1992nu}. 

\begin{itemize}
\item \textbf{Complex manifold}   \\
Intuitively, complex manifolds are topological spaces that locally look like flat complex space $\mathbb{C}^n$ for some $n$. More precisely, \\

 \textbf{Definition:}   Consider a real $2n$ dimensional manifold $M$. Then there is an atlas $\lbrace U_i, \psi_i \rbrace$ of open sets (which cover the manifold), and local coordinates. If we can ``complexify" the local coordinates, which means finding  homomorphisms $\psi_i : U_i \rightarrow \mathbb{C}^n$,  such that for any (non-empty) $U_i \cap U_j$,   $\psi_i o \psi_j^{-1} : \psi_j(U_i \cap U_j) \rightarrow \psi_i(U_i \cap U_j)$ is a holomorphic map from $\mathbb{C}^n$ to itself, then $M$ is called a complex manifold of dimension $n$.

In order to give a necessary and sufficient condition for when a real manifold is complex, one first defines an almost complex structure $J$ which is a $(n,n)$- tensor on $M$ (consider $M$ as a real manifold) such that $J^2=- 1$. This means it's possible to define the local complex coordinates. More concretely, choose a patch $U$ we have $2n$ real coordinates $\lbrace x_1,\dots x_n, y_1 \dots y_n\rbrace$, then $J$ acts on coordinate basis as 
\begin{equation}
J (\partial_{x_i})=\partial_{y_i},\qquad J(\partial_{y_i})=-\partial_{x_i}.
\end{equation}  

So by defining local complex coordinates,

\begin{equation}
z_j=x_j+ i y_j,\qquad \bar{z_j}= x_j- i y_j
\end{equation}
 we  get $J \partial_{z_i}=i \partial_{z_i}$, $J \partial_{\bar{z_i}}=-i \partial_{\bar{z_i}}$. Then $M$ being a complex manifold is equivalent to being able define complex coordinates in each patch such that under coordinate transformations almost complex structure stays diagonal (integrability). In this situation $J$ is called a complex structure tensor.
 
 The necessary and sufficient condition for $J$ to be complex structure is that the following tensor becomes zero (See \cite{Nakahara:1990th} Theorem 8.12 for a proof),
 
 \begin{equation}\label{Nijenhuis}
 N(v,w)= [v,w]+J[v,Jw]+J[jv,w]-[Jv,Jw],
 \end{equation}
where $v$,$w$ are arbitrary vector fields.  This is called $Nijenhuis$ tensor. 

\item \textbf{Intersection numbers:}\\
 As it is clear from the name it is the number of intersection points between cycles in $M$, so by Poincare duality we may be able to express the intersection number of divisors as the integration of the corresponding dual $(1,1)$-forms.  For example, consider $\mathbb{P}^n$. All of the divisors in this space can be written as $mH$, where $H$ is the hyperplane divisor corresponding to the vanishing locus of any linear polynomial. Then the intersection number of $n$ different divisors,
\begin{equation}
[m_1 H]\cdot [m_2 H] \cdots [m_n H]
\end{equation} 

can be written as the integral

\begin{equation}
\int_{\mathbb{P}^n} (m_1 \omega)\wedge (m_2 \omega) \wedge \cdots \wedge (m_n \omega)=m_1 m_2 \cdots m_n Vol (\mathbb{P}^n)
\end{equation}

where $\omega$ is the Kahler form of the projective space. So if we normalize the integral so that $\int J^n=1$, then we can use the integral above to say the intersection number is $m_1 m_2 \cdots m_n$. As another example, consider the product of two projective spaces $\mathbb{P}^{n_1}\times \mathbb{P}^{n_2}$. Similar to previous case we normalize the integral $\int \omega_1^{n_1} \omega_2^{n_2} =1 $ where $\omega_1$ and $\omega_2$ are Kahler forms of the two projective spaces. Then the intersection numbers can be computed as 

\begin{equation}
[m_1 H_1]\cdot [m_2 H_1] \cdots [m_{n_1} H_1]\cdot [l_1H_2] \cdots [l_{n_2} H_2]=m_1 \cdots m_{n_1} \cdot l_1 \cdots l_{n_2} \int_{\mathbb{P}^{n_1}\times \mathbb{P}^{n_2}} \omega_1^{n_1} \wedge \omega_2^{n_2} 
\end{equation} 
where $H_1$, and $H_2$ are the hyperplane divisors. For a general toric variety it's possible to figure out the intersection numbers from the toric data. The reader can refer to \cite{toric} for more information.

\item \textbf{Blow Up} \\

In this subsection we try illustrate the process described in Section \ref{manifold_building} by a simple example (See \cite{Hartshorne} I.4 , also II.7 for more abstract definitions). Consider $\mathbb{P}^2$ with homogeneous coordinates $(x,y,z)$. We choose the patch $z=1$, and consider the following hyperserface inside $\mathbb{P}^2 \times \mathbb{P}^1$,

\begin{equation}\label{P2blowup}
x u_1 - y u_2=0,
\end{equation} 

where $u_1$ and $u_2$ are the homogeneous coordinates of $\mathbb{P}^1$. We see from this equation whenever $(x,y)\neq (0,0)$, a single point in $\mathbb{P}^1$ is fixed, however when $(x,y)=(0,0)$, there is no constraint on $u_1$ and $u_2$. So we see that \ref{P2blowup} correspond to surface which generically seems to be the same as the original $\mathbb{P}^2$ plane, but the origin is replaced by a whole $\mathbb{P}^1$.

 This $\mathbb{P}^1$ is called an \textbf{exceptional divisor} $E$, and it can be shown since we've blown up a generic smooth point in $\mathbb{P}^2$, it's self intersection is $-1$,
 \begin{equation}
 E.E=-1.
 \end{equation}
This exceptional divisor can also be seen as the projectivization of the normal bundle to the point $(0,0,1)$  i.e. origin . \\

To see how blow up can be used to ``smooth out" the singularities consider a curve with double point singularity (node, or cusp) at the origin, as an example (in patch $z=1$),
\begin{equation}
y^2=-x^2(x-1),
\end{equation}
In this case we have a node (see Fig[\ref{BlownUpCurve}]). Now we rewrite $x=e x'$, and $y=e y'$, then it's clear from (\ref{P2blowup}) that $e=0$ is the locus of the exceptional divisor. Then the above equation becomes,
 
 \begin{equation}\label{blownupcurve}
 e^2 (y'^2+x'^2(e x'-1))=0,
 \end{equation}
the order $2$ zero at $e=0$ indicates the order of singularity. so if we remove this factor from the equation the rest of that will be a smooth curve.

\begin{figure}[h]
\begin{centering}
\includegraphics[width=0.9\textwidth]{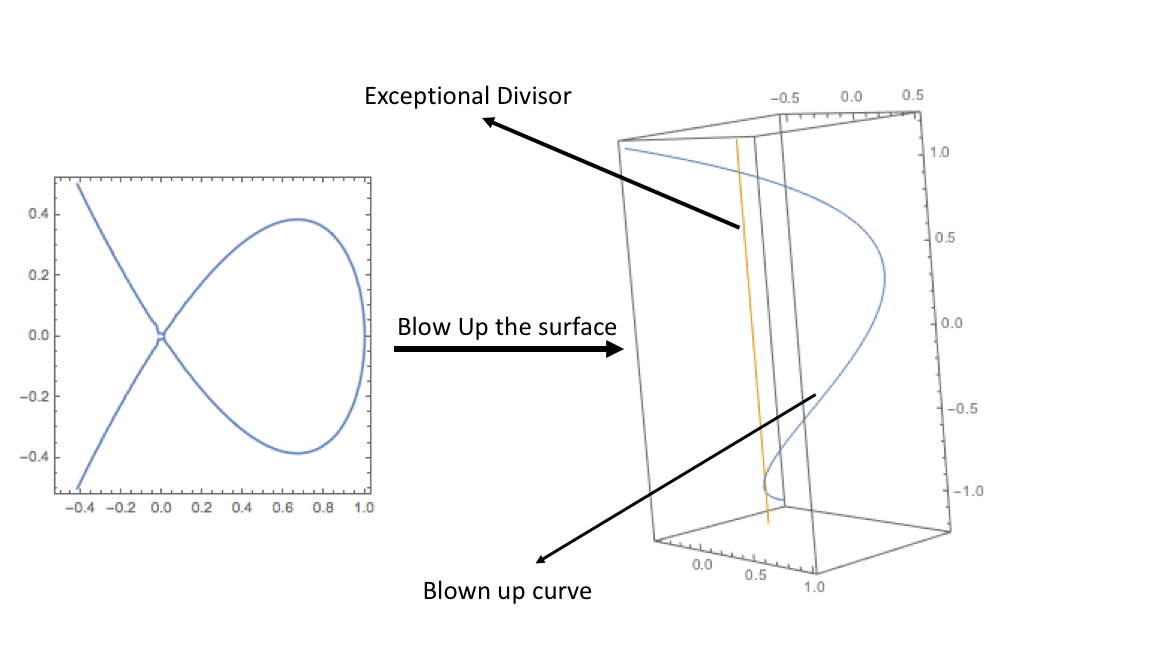}
\caption{\emph{The curve on the left correspond to the curve $y^2+x^2(x-1)=0$ in $\mathbb{P}^2$ inside the patch $z=1$, after adding a exceptional divisor in the origin, we get a reducible curve, one component correspond to the exceptional divisor (the $e^2$ factor above) shown as a orange line, and the other irreducible component is called the strict transform of the original curve on right curve. Note if we look at the curve on the right from top (or shrink the exceptional divisor to zero), its image over the horizontal plane will be the same as the curve on the left.}}\label{BlownUpCurve}
\end{centering}
\end{figure}

More formally we can describe what we did as a morphism,

\begin{equation}
\rho : \tilde{\mathbb{P}^2} \rightarrow \mathbb{P}^2,
\end{equation}

where $\tilde{\mathbb{P}^2}$ is the blown up projective plane (\ref{P2blowup}) (this is just the first Hirzebruch surface $\mathbb{F}_1$). Then the strict transformation of the curve will be,

\begin{equation}
\tilde{C}=\rho^* C -2 E,
\end{equation}

where $C$ is just the divisor class of the curve in the projective plane. The factor $2$ represents the double point singularity.
\\

\item \textbf{Vector Bundles} \\
\begin{itemize}
\item \textbf{Definition} Consider a compact manifold $M$ (real or complex), we can cover it with open sets and local coordinates $\lbrace U_i,\psi_i \rbrace$. Intuitively, a vector bundle locally looks like a product $U_i \times W$ where $W$ is a vector space with fixed dimension. To get a non trivial vector bundle over $M$, we need to glue these local structures. 
 
Again we need to define this more precisely. A vector bundle is given by a projection,

\begin{equation}
\pi : V \rightarrow M, 
\end{equation}  
where $V$ is the total space of the bundle, $M$ is the base manifold, and $\pi^{-1}(x) \sim W$ for any point $x$ in the base manifold. Similar to the definition of manifolds, there are homomorphisms(called local trivializations) $\phi_i : V \rightarrow U_i \times W=\pi^{-1}(U_i)$, and similar to coordinate transformation between patches, we need to define "transition functions" on $U_i\cap U_j$ as $t_{ij} =\phi_i o \phi_j^{-1} : U_j\times W \rightarrow U_i \times W$. Over any point $x\in U_i \cap U_j$, $t_{ij}(x)$ is just a homomorphism inside the vector space $W$. In principle the transition functions can be elements of Lie groups $G$ in various representations. $G$ is called the structure group of the bundle, and rank of the bundle $rk(V)$ is the dimension of $W$.

\item \textbf{Section} Sections are defined as maps $S : M\rightarrow V$. Locally this means over each open patch in the base manifold there is a map $S_i : U_i \rightarrow W$ such that for any $x \in U_i$, $S_i(x)$ is a unique vector in $W$. These local maps then glue together by the transition functions as $S_i =t_{ij}S_j$ to make a global section.

\item \textbf{Connection and Curvature} Similar to the tangent vectors of a manifold, we can define the parallel transport of elements in the vector bundle. To do this consider a local frame over $U_i$ (a basis of the vector space in $U_i \times W$ ) $\lbrace e_1 \dots e_p \rbrace $, then parallel transport of $e_i$ in direction $\mu$ in the base manifold is given defined by the connections: 

\begin{equation}
\nabla_{\mu} e_i = A_{\mu i}^j e_j.
\end{equation}

Note that $A_{\mu}$ is a one form with values in (adjoint representation of) the structure group. Also it's clear that under the ``local" transformations $e'_i=g(x)_{ji} e_j$, the connections transform in the following non-covariant way,

\begin{equation}
A'=g^{-1}A g+g^{-1}dg.
\end{equation}

The corresponding curvature, which is covariant under these transformations, is defined similar to curvature of manifolds,

\begin{equation}
F_{\mu \nu}=[\nabla_{\mu},\nabla_{\nu}],\qquad F_{\mu \nu} = \partial_{\mu} A_{\nu}-\partial_{\nu}A_{\mu}+[A_{\mu},A_{\nu}]
\end{equation}

\item \textbf{Gauge theory} There is a clear correspondence between physical gauge theories and vector bundles.  Structure groups, vector bundles, connections, curvature, and the transformations $g$, correspond to the gauge group, matter field, gauge fields, field strength and gauge transformations respectively.

\item \textbf{Holomorphic bundles} Suppose $\pi : V\rightarrow M $ is a complex vector bundle (which means the fibers are isomorphic to a $\mathbb{C}$-linear space) over a complex manifold. Then $V$ is called holomorphic if the transition functions are holomorphic relative to the complex coordinates. It can be shown for every holomorphic bundles, we can choose a gauge such that $A_{\bar{a}}$ components of the connection becomes zero. In other words $\nabla_{\bar{a}}=\partial_{\bar{a}}$. Also (if we can define a hermitian inner product on the fibers of $V$) the $(2,0)$ and $(0,2)$ components of the field strength are zero for holomorphic bundles (See \cite{Huybrechts} section 4.3 and appendix 4.B, also \cite{GSW2} 15.6 for more intuitive/physical discussion), 

\begin{equation}
F_{ab}=F_{\bar{a}\bar{b}}=0.
\end{equation}

\end{itemize}

If $V_1$ and $V_2$ are two bundles with structure groups $G_1$ and $G_2$, we can define the direct sum and direct product of bundles as $G_1\oplus G_2$ and $G_1 \otimes G_2$ structure groups respectively.

\item \textbf{Cohomology} There are various ways to define cohomology groups. We only briefly discuss the de Rahm and Dolbeault cohomology here (See \cite{Hartshorne} III.4 for Cech cohomology). We will consider the case of complex, compact, K\"ahler manifolds.

Generally when there is a complex as,

\begin{equation}
0 \rightarrow A^0 \xrightarrow{d^0} A^1 \xrightarrow{d^1} A^2 \xrightarrow{d^2} \dots
\end{equation}
 such that $d^{i+1} \circ d^i=0$, then cohomology groups are defined as,
 \begin{equation}
 H^i =\frac{Ker\left(d^i : A^i \rightarrow A^{i+1}\right) }{Im\left(d^{i-1} : A^{i-1} \rightarrow A^{i}\right)}
 \end{equation}

\begin{itemize}

\item \textbf{de Rahm Cohmology}

One example of cohomology is the de Rahm cohomology over a real manifold $M$ defined by the differential operator $d$ on a complex of differential forms,

\begin{equation}
0 \rightarrow \Omega^0 \xrightarrow{d} \Omega^1 \xrightarrow{d} \Omega^2 \xrightarrow{d} \dots
\end{equation}  
 
where $\Omega^n$ is the space of $n$-forms, and the corresponding cohomology groups  $H^n(X,\mathbb{R})$ are the space of closed $n$-forms modulo the exact ones. The dimension of these groups are called Betti numbers $b_n$

\item \textbf{Dolbeault Cohomology}

On a complex manifold we can decompose the differential operators into holomorphic and anti-holomorphic parts $d=\partial+\bar{\partial}$, where $\partial^2=\bar{\partial}^2=0$, and correspondingly, the decompose into the direct sum of mixed $(p,q)$-forms,

\begin{equation}
\Omega^n = \bigoplus_{p+q=n} \Omega^{(p,q)},
\end{equation} 
 
 where elements of $\Omega^{(p,q)}$ can be written as $\omega_{a_1\dots a_p \bar{a}_1 \dots \bar{a}_q} dz^{a_1}\dots dz^{a_p} d\bar{z}^{\bar{a_1}}\dots d\bar{z}^{\bar{a_q}}$. Since $\bar{\partial}^2=0$, we can define the Dolbeault cohomology relative to $\bar{\partial}$,
 \begin{equation}
 0 \rightarrow \Omega^{(p,0)} \xrightarrow{\bar{\partial}} \Omega^{(p,1)} \xrightarrow{\bar{\partial}} \Omega^{(p,2)} \xrightarrow{\bar{\partial}} \dots.
 \end{equation}

 \begin{equation}
 H^{p,q}_{\bar{\partial}}(X) :=H^q_{\bar{\partial}} (X,\Omega^{(p,0)}) =  \frac{Ker\left(\bar{\partial}: \Omega^{(p,q)}\rightarrow \Omega^{(p,q+1)} \right)}{Im\left(\bar{\partial}: \Omega^{(p,q-1)}\rightarrow \Omega^{(p,q)} \right)}.
 \end{equation}

The dimension of the cohomology group $H^{p,q}_{\bar{\partial}}(X)$ is called the \textbf{Hodge number} $h^{p,q}$. If the manifold $M$ is also compact, we get the following relations, 
 \begin{equation}
 b_n = \sum_{p+q=n} h^{p,q} 
 \end{equation}

Similarly on a holomorphic vector bundle $V$, the connection also decomposes (by complexity of the bundle) $\nabla = \nabla^{(1,0)}+\nabla^{(0,1)}$, and by holomorphicity condition, $(\nabla^{(1,0)})^2=(\nabla^{(0,1)})^2=0$. So we may define cohomology groups $H^n(X,V)$ with respect to the differential operator $\bar{\nabla}=\nabla^{(0,1)}$. The elements $\psi^x_{\bar{a}_1\dots\bar{a}_n}$ of this group are $(0,n)$-forms with values in $V$  the upper index $x$ is the vector bundle index which correspond to a representation of the structure (gauge) group. Also these element are $\nabla^{(0,1)}$ closed but not exact. 

To get a little physical intuition consider $n$-forms living in the internal compact manifold with an index transforming in some representation of the gauge group. One way to see how these fields arise in string theory is from the fact that the space of zero modes og gauginos decompose into subspaces isomorphic to the space of differential $n$-forms. Since they are zero modes, they must be $\bar{\nabla}$-closed. However $\bar{\nabla}^2=0$, so we always have a gauge freedom,

\begin{equation}
\psi^x \rightarrow \psi^x +( \bar{\nabla} \Lambda)^x,
\end{equation}

where $\Lambda$ is an arbitrary $n-1$-form. Since the theory is invariant under this ``gauge transformation", it justifies to consider only the elements of the cohomology groups as the space of physical solution of the massless Dirac equation (these are discussed in detail in \cite{GSW2} chapters 13 to 16).  

\end{itemize}

\item \textbf{Chern classes}\\

Characteristic classes (including Chern classes) are elements of the cohomology groups that are invariant under smooth deformations, and measure the ``non-triviality" of the bundles. There are various ways to define the Chern classes, here we use the differential geometric definition. Here we restrict ourselves to the complex vector bundles $\pi : V\rightarrow M$ with curvature $2$-form $F$, and rank $n$. 

\begin{itemize}
\item \textbf{Chern class} 

The total Chern class is defined as
\begin{equation}
c(V) = det(1+i\frac{F}{2\pi})
\end{equation}

We can expand this order by order to get,

\begin{eqnarray}
c (V) &=& 1+c_1(V)+c_2(V)+\dots,\\ 
c_1(V) &=&  \frac{i}{2\pi} (tr F), \\
c_2(V) &=& \frac{1}{2} \left(\frac{i}{2\pi} \right)^2 \left( tr(F \wedge F) - tr(F)\wedge tr(F) \right), \\
\vdots \nonumber \\
c_n (V) &=& \left( \frac{i}{2\pi}\right)^n det(F)
\end{eqnarray}
\item \textbf{Chern character}

\begin{eqnarray}
ch(V) &=& tr\left(e^{i\frac{F}{2\pi}}\right), \\
ch_0(V) &=& n, \\
ch_1(V) &=& i\frac{F}{2\pi} = c_1(V), \\
ch_2(V) &=& -\frac{1}{4\pi^2} tr \left( F\wedge F \right) = \frac{1}{2} \left(c_1(V)^2-2c_2(V) \right), \\
\dots
\end{eqnarray}

\item \textbf{Properties}

There are important identities for Chern classes/characters of direct sum and direct product of vector bundles. One can figure out these identities by trying to understand what is the corresponding curvature $2$-form, 
\begin{eqnarray}
V &=& V_1 \oplus V_2, \nonumber\\
F_V &=&\left( \begin{array}{cc}
F_1 & 0 \\
0 & F_2
\end{array}\right), \\
W &=& V_1 \otimes V_2, \nonumber \\
F_w &=& F_1 \otimes 1 + 1\otimes F_2
\end{eqnarray}

Then by the definition, the following identities hold,

\begin{eqnarray}
c(V) &=& c(V_1)\wedge c(V_2), \\
ch(V) &=& ch(V_1)+ch(V_2), \\
ch(W) &=& ch(V_1)\wedge ch(V_2).
\end{eqnarray}
The first two relations will be true even for non-trivial extensions,
\begin{equation}
0\rightarrow V_1 \rightarrow V \rightarrow V_2 \rightarrow 0.
\end{equation}

\item \textbf{Atiyah-Singer index theorem}
Suppose $V$ is a holomorphic vector bundle over a compact complex manifold $M$ of complex dimension $m$, then the \textbf{Euler Characteristic} of $V$ is defined as
\begin{equation}
\chi (M,V) = \sum_{i=1}^m (-1)^m h^m(M,V).
\end{equation}

The following theorem connects the Euler characteristic and Chern classes,

\begin{equation}
\chi (M,V) = \int_M ch(V) td(M)
\end{equation}
where $td(M)$ is the total Todd class, and it's relation with Chern classes of the tangent bundle $TM$ is given as,

\begin{eqnarray}
td_0(M)&=&1, \\
td_1(M)&=&\frac{1}{2} c_1(M), \nonumber\\
td_2(M)&=& \frac{1}{12 }\left(c_1(M)^2+c_2(M)\right), \nonumber \\
\dots \nonumber
\end{eqnarray} 
This theorem is important because it gives the chirality of the effective field theories in terms of the topological quantities of the extra dimensional manifold and the gauge bundles over it.

\item \textbf{Stability of $V$} 

Another important quantity that is defined for complex vector bundles over compact Kahler manifolds of complex dimension $m$ is the slope of bundle,

\begin{equation}
\mu(V) = \frac{1}{rank(V)} \int c_1(V) \wedge \omega^{m-1}
\end{equation}
where $\omega$ is the Kahler class of the complex manifold. 

\textbf{Definition} A holomorphic vector bundle $V$ over a compact Kahler manifold is called slope (Mumford) stable is for every sub-sheaf $\mathcal{F}\subset V$, $\mu(\mathcal{F}) < \mu(V)$ 

We have seen that in string theory compactification $4$-dimensional supersymmetry puts the following constraint on the holomorphic bundles,

\begin{equation}
g^{a\bar{b}}F_{a\bar{b}}=0.
\end{equation}

A theorem by Donaldson-Uhlenbeck-Yau \cite{UY} \cite{Donaldson}, states that for any holomolrphic vector bundle over a compact Kahler manifold, the above condition is satisfied if and only if the bundle is poly-stable\footnote{To be more precise, the connection of the vector bundle must satisfy $g^{a\bar{b}}F_{a\bar{b}}=\frac{-i}{Vol(M)} \mu(V) 1$ if and only is it's stable. However in heterotic compactifications we are restricted to the case of zero first Chern class, and therefore zero slope (see e.g. \cite{Huybrechts} Ch. 4.B). }. A poly-stable bundle is simply a direct sum of stable bundles, all with the same slope: $V=\bigoplus_i V_i$ with $\mu(V_i)=\mu(V)$ $\forall i$.
\end{itemize}

\end{itemize}

\end{document}